\documentclass[aps,prc,amsmath,showpacs,superscriptaddress,nofootinbib]{revtex4}
\usepackage{graphicx,epsfig,epstopdf,longtable}
\newcommand{\beqy}{\begin{eqnarray}}
\newcommand{\eeqy}{\end{eqnarray}}
\newcommand{\bmlet}{\begin{subequations}}
\newcommand{\emlet}{\end{subequations}}
\newcommand{\bfdel}{\pmb{\nabla}}
\begin{document}

\def\gsimeq{\,\,\raise0.14em\hbox{$>$}\kern-0.76em\lower0.28em\hbox
{$\sim$}\,\,}
\def\lsimeq{\,\,\raise0.14em\hbox{$<$}\kern-0.76em\lower0.28em\hbox
{$\sim$}\,\,}

\title{Role of neutron pairing with density-gradient dependence in the
semi-microscopic treatment of the inner crust of neutron stars.}

\author{N. Chamel}
\affiliation{Institut d'Astronomie et d'Astrophysique, CP-226, Universit\'e
Libre de Bruxelles, 1050 Brussels, Belgium}

\author{J. M. Pearson}
\affiliation{D\'ept. de Physique, Universit\'e de Montr\'eal, Montr\'eal
(Qu\'ebec), H3C 3J7 Canada}

\author{N. N. Shchechilin}
\affiliation{Institut d'Astronomie et d'Astrophysique, CP-226, Universit\'e
Libre de Bruxelles, 1050 Brussels, Belgium}

\date{\today}

\begin{abstract}
Using the fourth-order extended Thomas-Fermi method with Strutinsky-integral 
shell and pairing corrections,  we calculate the inner crust of neutron 
stars with the BSk31 functional, whose pairing has two terms: i) a term that is 
fitted to the results of 
microscopic calculations on homogeneous nuclear matter (accounting for both 
medium polarization and self-energy effects) that are more realistic 
than those of our earlier functionals; ii) an empirical term that is
dependent on the density gradient, which permits an excellent fit to nuclear 
masses. Both proton and neutron pairing are taken into account, the former 
in the BCS theory and the latter in the local density approximation. We found 
that the equilibrium value of the proton number $Z$ remains 40 over the entire 
density range considered, whether or not neutron pairing is included.
The new equation of state and the composition are very similar to those of our 
previously preferred functional, BSk24. However, the predicted neutron 
pairing fields are quite different. 
In particular, clusters are found to be impermeable to the neutron superfluid. 
The implications for the neutron superfluid dynamics are briefly discussed. 
Since the new pairing is more realistic, 
the functional BSk31 is better suited for investigating neutron superfluidity 
in neutron-star crusts. 
\end{abstract}

\maketitle

\section{Introduction}
\label{intro}

Three distinct regions are conventionally recognized in neutron stars (see, for
example, Ref.~\cite{bc18} for a recent review). The outermost of these regions,
the ``outer crust'', consists of a Coulomb crystal of bound nuclei and electrons that 
globally is electrically neutral. The nuclei in this region become more and 
more neutron-rich with increasing depth, until at a mean baryon number density 
$\bar{n}$ of around $2.6 \times 10^{-4}$ fm$^{-3}$ unbound neutrons start to 
appear. This so-called ``neutron drip'' marks the transition to the ``inner 
crust'', an inhomogeneous assembly of neutron-proton clusters arranged on a 
crystal lattice and unbound neutrons, with electrons assuring global
neutrality. By the point where $\bar{n}$ has risen to about 0.08 fm$^{-3}$ the 
inhomogeneities have been smoothed out: the ``core'' of the star has been 
reached.

In our 2018 paper~\cite{pea18}, the inner crust was calculated in the framework
of spherical Wigner-Seitz (WS) cells, with both droplet and bubble 
configurations being admitted. To calculate the energy per nucleon in this 
framework, we used the ETFSI+pairing (fourth-order extended Thomas-Fermi plus
Strutinsky integral with pairing) approach, a high-speed approximation to the 
Hartree-Fock- Bogoliubov (HFB) method consisting of two distinct stages: a 
semi-classical extended Thomas-Fermi (ETF) calculation of the total energy, 
followed by the addition of proton shell corrections, calculated by the 
Strutinsky-integral method (SI), and proton pairing corrections, handled in the
Bardeen-Cooper-Schrieffer (BCS) approximation. For these calculations, we 
used a family of nuclear energy density functionals that we developed not 
only for the study of neutron-star structure but also for the general purpose 
of providing a unified treatment of a wide variety of phenomena associated with
the birth and death of neutron stars, such as core-collapse supernovae 
and neutron-star mergers, along with the r-process of nucleosynthesis 
(see Ref.~\cite{gcp13} and references therein). These functionals are based on 
generalized Skyrme-type forces and density-dependent contact pairing forces, 
the parameters of which were determined primarily by fitting to essentially all
the nuclear-mass data of the 2012 Atomic Mass Evaluation~\cite{ame12}. In those
fits we calculated nuclear masses using the HFB method with axially-symmetric 
deformation taken into account; the necessary HFB formalism for our generalized
Skyrme functionals is given in Refs.~\cite{cgp09,gcp09,cha10}. Our fits were 
made subject to certain constraints, the most significant of which was to 
require consistency, up to the densities prevailing in neutron-star cores, with 
the equation of state (EoS) of homogeneous pure neutron matter, as calculated 
{\it ab initio} from realistic two- and three-nucleon forces.
Later, we showed~\cite{pea20,pea22,shch23}, that at densities approaching that 
of the crust-core transition configurations with non-spherical ``pasta" shapes 
became energetically favored. ``Spaghetti" and ``lasagna" shapes were first 
predicted by Ravenhall {\it et al.}~\cite{rav} and Hashimoto 
{\it et al.}~\cite{hash} some 
forty years ago, and have subsequently been found in many different approaches. 
This part of the neutron star is known as the "mantle". While it is often considered 
to be an integral part of the inner crust there are good reasons for treating it as a 
distinct fourth region, since it is expected to behave like liquid crystals~\cite{pot98}.  
Neither in Ref.~\cite{pea18} nor in any other of our papers on the EoS did we 
include SI or pairing corrections for neutrons. The omission of the SI 
correction for
neutrons was justified by the argument that neutron shell effects are known to 
be much smaller than proton shell effects~\cite{oy94}; in fact, we showed 
in Section I of Ref.~\cite{pea22} that the situation is a little more 
complicated, but that the omission of the neutron SI correction is nevertheless
expected to be a fairly good approximation. 
Indeed, a comparison of full 3D band-theory calculations with HF calculations 
performed in spherical WS cells shows that the neutron shell effects found in 
the latter are largely spurious: they arise as a result of the discretization 
imposed by the spherical WS cell approximation on the quasi-continuous spectrum
of the unbound single-particle (s.p.) neutron states~\cite{ch07}. Thus in this 
respect the ETFSI+pairing method may actually be more reliable than HFB 
calculations performed in spherical WS cells. 

On the other hand, we certainly need to take neutron pairing into account if
we wish to include neutron superfluidity in the list of phenomena that our
functionals can describe. All the functionals used in  Ref.~\cite{pea18}, 
BSk22, BSk24, BSk25 and BSk26,
contain a neutron-pairing term, and it would have been possible to take
account of it in the EoS. As a matter of fact, neutron pairing for BSk22 and
BSk24 has recently been included within the ETFSI framework in 
Refs.~\cite{sp20,sp21} in the local-density 
approximation (LDA). Moreover, $^1S_0$ pairing gaps and non-dissipative 
mutual entrainment coefficients of the neutron-proton superfluid mixture in the
outer core of a neutron star have been calculated for BSk24 in 
Ref.~\cite{alc21}, varying the temperature and the superflow velocities. 
However, such calculations are somewhat incomplete for the following reason. 
The pairing term in the functionals of 
Ref.~\cite{pea18} is determined analytically at each point in the inhomogeneous
nuclear system in question (nucleus in the original HFB fits, WS cell in the 
inner crust of neutron stars) in such a way as to reproduce the $^1S_0$
pairing gaps of homogeneous nuclear matter (HNM) of the appropriate density and
charge asymmetry, as determined in the {\it ab initio} calculations made by Cao
et al.~\cite{cao06} with realistic two- and three-nucleon forces. These latter
gap calculations were made both with and without self-energy
corrections~\cite{cao06}, but the excellent mass fits found for the functionals
of Ref.~\cite{pea18} were obtained by adopting the latter option, since taking 
account of the self-energy corrections led to gaps that are much smaller, too 
small to obtain good mass fits. However, while good mass fits are essential for
a reliable determination of the composition of the crust, gaps calculated with 
self-energy corrections are more realistic and thus more appropriate for the 
reliable study of superfluidity and superconductivity in neutron stars.

It was the realization of this point that led to the construction of a new 
family of functionals, BSk30--32~\cite{gcp16}, functionals that achieve the dual
purpose of retaining the excellent mass fits of the earlier functionals while 
simultaneously being much more suitable for the calculation of superfluidity
and superconductivity in neutron stars. 
They do this by having two pairing terms, the first of 
which is fitted to the HNM gaps of Ref.~\cite{cao06} calculated {\it with} the 
self-energy corrections included, to which is added a phenomenological surface 
term, i.e., one dependent on the local density gradients, the strength of which
is a parameter of the mass fit. The three functionals BSk30--32 were fitted to 
HNM symmetry coefficients of $J$ = 30, 31 and 32 MeV, respectively, and BSk31 
gave the best mass fit of the three, with a root mean square (rms) error of 
0.571 MeV for 2353 measured masses, as compared with 0.549 MeV for BSk24, our 
previously preferred functional, whose pairing is less realistic. The overall 
mass fit of BSk30 is almost as good, but BSk31 does significantly better for 
neutron-rich nuclei. 
An examination of the errors of the different functionals suggests that the 
best fit would have been obtained with a value of $J$ of around 30.6 MeV,
midway between the values corresponding to BSk30 and BSk31, and close to the 
value for BSk24, 30 MeV.

The present paper describes the first EoS calculations made with the functional
BSk31, and is the first in which we include neutron pairing. This functional is
just as well adapted to a unified treatment of all
three regions of neutron stars as the functionals we used in Ref.~\cite{pea18},
and we intend to make such calculations in the near future. However, in this
paper we confine ourselves to the inner crust, dealing with the special 
problems posed by the density-gradient dependence of the pairing. Moreover, 
since our handling of
pasta is undergoing further refinement we will limit our calculations in the 
present paper to densities lower than 0.06 fm$^{-3}$ so that the WS cells can 
safely be assumed to be spherical; this restriction will be of no consequence 
for our main concern here, which is to demonstrate the role of neutron pairing,
and the changes that it brings about. The fact that the proton pairing is 
also more realistic with functional BSk31 means that more reliable calculations
of proton superconductivity will be possible, but this phenomenon only comes 
into play at densities higher than 0.06 fm$^{-3}$, 
where protons become unbound, and will not be considered here. 

Our calculation of the properties of the inner crust with the functional BSk31
are made with our usual ETFSI method. The new feature, neutron pairing, is
handled in  the LDA; for proton pairing, however, we retain the usual 
BCS method~\cite{pcpg15}. Using functionals BSk24 and SLy4, the authors of 
Ref.~\cite{sp20} find that provided the ETFSI method takes neutron 
pairing into account it is in good agreement, as far as the energy per nucleon 
is concerned, with HFB calculations performed in spherical WS cells. The only 
difference is that these HFB calculations 
show the optimum value of the number $Z$ of protons in the spherical WS cell to be fluctuating 
between 36 and 50 as the mean density changes, while there are no such 
fluctuations with the ETFSI calculations. However, since the energy changes 
associated with the fluctuations are very small there is no impact on the level
of agreement between the two methods.
In any case, the fluctuations in $Z$ could be spurious, as explained above. 
We thus conclude that the ETFSI method used here provides an accurate 
approximation to the energy per nucleon obtained in 
the HFB method within the spherical WS cell approximation. Deviations in the 
optimum values of $Z$ lie within the errors of such implementation of the HFB 
method in the range of densities $\bar n$ considered here. Moreover, the high 
speed of the ETFSI approximation is of crucial importance for the large-scale 
calculations that we are undertaking. 

Section~\ref{method}, where we describe our method of calculation, is devoted
primarily to our handling of the new pairing, with its dependence on the
density gradient, while our results are presented and discussed in
Sections~\ref{results} and~\ref{concl}, respectively.

\section{Method of calculation}
\label{method}

Adopting the ETFSI method, we write the energy per nucleon as
\beqy\label{1}
e_\mathrm{ETFSI} = e_\mathrm{ETF} + 
\frac{1}{A}\left(E^\mathrm{SI}_p + 
E^\mathrm{pair}_p + E^\mathrm{pair}_n\right)  \quad  ,
\eeqy
in which $e_\mathrm{ETF}$ is the energy per nucleon calculated by the ETF 
method, $E^\mathrm{SI}_p$ is the SI shell correction for protons in a WS cell with $A$ nucleons and 
$E^\mathrm{pair}_q$ is the pairing energy for protons or neutrons, 
as $q$ = $p$ or $n$, respectively. The justification for the neglect of the SI
correction for neutrons has been discussed above in Section I. 

The ETF method consists of expanding the Bloch density matrix in powers of 
$\hbar$ \cite{bgh85}, so that the ETF energy becomes a functional of only 
the nucleon densities and their gradients. We parametrize the spherically 
symmetrical density distributions according to 
\beqy\label{2}
\widetilde{n_q}(r) = n_{\mathrm{B}q} + \frac{n_{\Lambda q}}{1 + \exp \left[\Big(\frac{C_q - R}
{r - R}\Big)^2 - 1\right] \exp \Big(\frac{r-C_q}{a_q}\Big) } \quad ,
\eeqy
in which the first term represents a constant background and the second describes the 
cluster centered around $r=0$. 
Here $n_{\Lambda q}$ modulates the density excess due to the cluster, while the
geometrical parameters $C_q$ and $a_q$ control the cluster size and the 
diffuseness of its surface. The first exponential factor in the denominator of 
the second term of Eq.~(\ref{2}) was introduced~\cite{ons08} in order for the first derivative of 
the density profile to vanish on the cell surface at $r=R$, a necessary condition 
established by Wigner and Seitz~\cite{ws33} (note that this condition is not 
satisfied by the simple Fermi parametrization adopted in 
Refs.~\cite{sp20,sp21}). In fact, with this profile {\it all} derivatives
vanish on the cell surface, which allows us to use an integrated form of the
fourth-order ETF method, in which only first- and second-order derivatives of 
the density appear. It was recently shown that the resulting profile is not 
suitable for pasta, but is quite acceptable for densities below 0.06 fm$^{-3}$, to which we limit
ourselves here.~\cite{shch24}. 
For a fixed value of $Z$, the number of protons per cell, the ETF energy per 
nucleon is minimized with respect to the geometrical parameters of the neutron 
and proton distributions~(\ref{2}) and $N$, the number of neutron per cell (not
necessarily an integer, since some neutrons are unbound and thus not confined
to one cell). 

With the ETF part of the calculation complete, our code then computes 
the smooth single-particle (s.p.) proton central and spin-orbit fields, 
$\widetilde{U_p}(r)$ and $\widetilde{\pmb{W_p}}(r)$, respectively, the Coulomb 
field $\widetilde{U_\mathrm{C}}(r)$ and the effective proton mass
$\widetilde{M^*_p}(r)$.
The HF equation 
\beqy
\left\{-\pmb{\nabla}\frac{\hbar^2}{2\widetilde{M^*_p}(r)}\cdot\pmb{\nabla} +
\widetilde{U_p}(r) + \widetilde{U_\mathrm{C}}(r) - 
i\widetilde{\pmb{W_p}}(r)\cdot\bfdel
\times\mbox{\boldmath$ \sigma$}\right\}\psi_{p,\nu}=
\widetilde{\epsilon}_{p,\nu}\psi_{p,\nu} \quad ,
\eeqy 
is then solved with these fixed fields 
for the s.p. proton energies $\widetilde{\epsilon}_{p,\nu}$, after which proton
pairing is calculated with our usual handling of the BCS method, as
in Refs.~\cite{pcpg15,pea18}. The proton SI correction appearing in 
Eq.~(\ref{1}) then becomes (see Appendix~\ref{app:Strutinsky} for the proof)
\beqy\label{4}
E^\mathrm{SI}_p = \sum_{\nu}V^2_{p,\nu}\widetilde{\epsilon}_{p,\nu} -
\int d^3\pmb{r}\Biggl\{\frac{\hbar^2}{2\widetilde{M^*_p}(r)}
\widetilde{\tau_p}(r) + \widetilde{n_p}(r)\left[\widetilde{U_p}(r)
+\widetilde{U_\mathrm{C}}(r) \right]
+\widetilde{\pmb{J_p}}(r)\cdot\widetilde{\pmb{W_p}}(r)\Bigg\}
\quad ,
\eeqy
in which the $V^2_{p,\nu}$ quantities are the s.p. occupation probabilities, as
given, for example, by Eq.~(7) of Ref.~\cite{pcpg15}; the summation goes over
all the s.p. proton states. Also $\widetilde{n_p}(r)$, $\widetilde{\tau_p}(r)$ 
and $\widetilde{\pmb{J_p}}(r)$ are the smoothed values of the proton density $n_p(r)$, 
the kinetic proton density $\tau_p(r)$ and the proton spin-current vector density 
$\pmb{J_p}(r)$ emerging from the minimization of the ETF energy in the 
first stage of the calculation; their presence in the second stage of the
ETFSI method ensures a high level of self-consistency in the calculation of
the shell and pairing corrections.    
 
For the proton pairing energy appearing in Eq.~(\ref{1}) we take the 
BCS expression
\beqy\label{5}
E^\mathrm{pair}_p = -\frac{1}{4}\sum_{\nu}\frac{\Delta^2_{p,\nu}}{E_{p,\nu}}
\quad ,
\eeqy
where the proton gaps $\Delta_{p,\nu}$ are given by the usual BCS equations, as
in Eq.~(9b) of Ref.~\cite{pcpg15}, and the proton quasi-particle energies 
$E_{p,\nu}$ by Eq.~(8) of Ref.~\cite{pcpg15} 
(see also Appendix~\ref{app:Strutinsky}).

The numerical implementation of the BCS method for neutrons is much more 
challenging since their unbound s.p. states form a quasi-continuum (see, e.g., 
Ref.~\cite{cgpo10}), and we adopt rather an LDA. In this way 
$E^\mathrm{pair}_n$ becomes a functional of the 
neutron and proton densities only,  and it is thus natural to treat it as being 
part of the ETF energy 
and therefore to optimize the ETF energy including the neutron pairing term. 
If this term were added after minimization, the equilibrium value of $N$
for given $Z$ would remain unchanged. This would certainly be wrong physically 
and would disagree with exact HFB calculations.

The neutron pairing energy $E_n^\mathrm{pair}$ is thus determined by adding 
to the energy density, {\it before} optimization of the ETF part of the 
calculation, the quantity~\cite{pizzo77}
\beqy \label{6}
{\mathcal E}_{\mathrm{cond},n}(r) =
-\frac{3}{8}\widetilde{n_n}(r)\frac{\widetilde{\Delta_n}(r)^2}{\epsilon_{Fn}(r)}
\quad  ,
\eeqy
where the neutron pairing field $\widetilde{\Delta_n}(r)$ 
is defined as the neutron pairing gap obtained locally by solving at each point $r$ 
in the inhomogeneous matter distribution the same 
BCS gap equations as in homogeneous nuclear matter with neutron density 
$\widetilde{n_n}(r)$ and proton density $\widetilde{n_p}(r)$ (see below).  
Here $\epsilon_{Fn}(r)$ is the local neutron Fermi energy,
\beqy \label{7}
\epsilon_{Fn}(r) =
\frac{\hbar^2 k_{Fn}(r)^2}{2\widetilde{M^*_n}(r)} \quad ,
\eeqy
where 
\beqy \label{7b}
k_{Fn}(r) = \left[3\pi^2\widetilde{n_n}(r)\right]^{1/3} \quad . 
\eeqy
Then
\beqy\label{8}
E_n^\mathrm{pair} =\int d^3\pmb{r}\, {\mathcal E}_{\mathrm{cond},n}(r) \quad .
\eeqy

The whole process is then repeated for different values of $Z$ in order
to achieve optimization with respect to this parameter also.
For further details on the ETFSI method, as we have used it in the past,
Refs.~\cite{ons08,pcgd12,pcpg15,pea18} should be consulted, since we devote 
the rest of this section to the density-gradient part of the pairing term of 
the BSk30-32 functionals, which we have to include in the ETFSI formalism for 
the first time. 

Our pairing interaction has the form, for two nucleons of charge type $q$ ($n$ or 
$p$) at positions $\pmb{r_i}$ and $\pmb{r_j}$ respectively 
(introducing the relative and center-of-mass coordinates $\pmb{r_{ij}}=\pmb{r_i}-\pmb{r_j}$ and 
$\pmb{r}=(\pmb{r_i}+\pmb{r_j})/2$ respectively) 
\beqy\label{P0}
v^{{\rm pair},q}(\pmb{r_i}, \pmb{r_j}) = 
f^{\pm}_q v^{\pi, q}(\pmb{r})~\delta(\pmb{r_{ij}}) \quad ,
\eeqy
\beqy\label{P1}
v^{\pi, q}(\pmb{r})= 
v^{\mathrm{hom}, q}[n_n(\pmb{r}),n_p(\pmb{r})]+\kappa_q |\nabla\,n(\pmb{r})|^2 \quad ,
\eeqy
which, with a slight change of notation, is just Eq.~(9) of Ref.~\cite{gcp16}.
The $f^{\pm}_q$ factors here are for fine tuning and are always very close to 
or exactly equal to unity, while $n = n_n + n_p$.
The $\kappa_q$ term, which was absent in the functionals used in our previous
calculations of the inner crust, was introduced to give a better fit to nuclear masses
while reproducing exactly the realistic HNM pairing calculations of Cao et al.~\cite{cao06}. 
In achieving this end, it must be presumed that our surface pairing has been improved.

All our functionals starting from BSk16~\cite{cgp08} and prior to those of 
Ref.~\cite{gcp16} (with the exception of BSk27~\cite{gcp13b}, which had an
older form of pairing) had only the
first term of Eq.~(\ref{P1}). Aside from the fine-tuning parameters 
$f^{\pm}_q$, the pairing in those earlier papers was completely determined 
{\it ab initio}, being given at any point $\pmb{r}$ where the neutron density 
is $n_n(\pmb{r})$ and the proton density is $n_p(\pmb{r})$ by~\cite{cha10}
\beqy \label{P2}
v^{\mathrm{hom}, q}[n_n(\pmb{r}),n_p(\pmb{r})]=-\frac{8\pi^2}{I_q[n_n(\pmb{r}),n_p(\pmb{r})]}
\left(\frac{\hbar^2}{2 M_q^*[n_n(\pmb{r}),n_p(\pmb{r})]}\right)^{3/2}   \quad .
\eeqy
Here $M_q^*[n_n(\pmb{r}),n_p(\pmb{r})]$ is the local effective mass for nucleons of charge type $q$, while
\beqy\label{P3}
I_q[n_n(\pmb{r}),n_p(\pmb{r})]=\sqrt{\epsilon_{Fq}(\pmb{r})}
\biggl[ 2\ln\left(\frac{2\epsilon_{Fq}(\pmb{r})}{\Delta_q(\pmb{r})}\right)+ 
\Lambda\left(\frac{\varepsilon_\Lambda}{\epsilon_{Fq}(\pmb{r})}\right) \biggr] \quad .
\eeqy
In this last expression $\Delta_q(\pmb{r}) = \Delta_q[n_n(\pmb{r}), n_p(\pmb{r})]$ is the 
local pairing gap for HNM with neutron density $n_n(\pmb{r})$ and 
proton density $n_p(\pmb{r})$ from the calculations of Cao et al.~\cite{cao06}, 
and $\epsilon_{Fn}(\pmb{r})$ is the local Fermi energy given by Eq.~(\ref{7}) 
for neutrons (the expression is similar for protons). It is to be noted that 
prior to Ref.~\cite{gcp16}, Eq.~\eqref{P2} and the local Fermi energy entering Eq.~\eqref{P3} were  
calculated replacing the effective mass by the bare mass for consistency with 
the absence of self-energy corrections in the adopted HNM gaps of Cao et 
al.~\cite{cao06} (these gaps were calculated using the free single-particle energy 
spectrum).  Also $\varepsilon_{\Lambda}$ is the cutoff above the Fermi level, while
\beqy \label{P5}
\Lambda(x)=\ln (16 x) + 2\sqrt{1+x}-2\ln\left(1+\sqrt{1+x}\right)-4 \quad .
\eeqy
Equation~\eqref{P2} is nothing but the BCS gap equation in homogeneous nuclear 
matter with neutron density $n_n(\pmb{r})$ and proton density $n_p(\pmb{r})$. Here, 
it was solved for the pairing interaction
$v^{\mathrm{hom}, q}[n_n(\pmb{r}),n_p(\pmb{r})]$ given the pairing gaps $\Delta_q(\pmb{r})$. 

\begin{figure}
\centerline{\includegraphics[width=0.7\columnwidth]{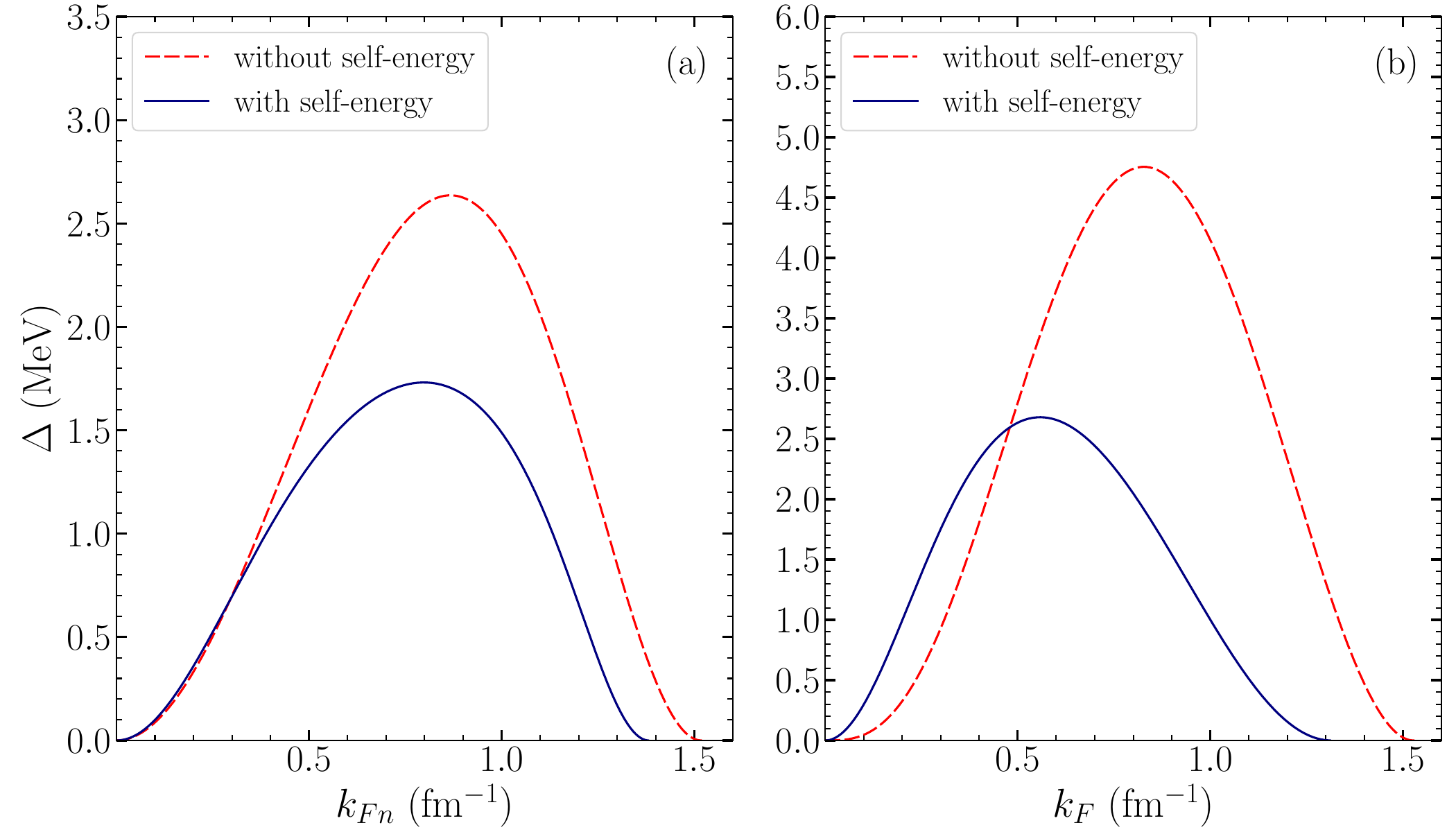}}
\caption{The $^1S_0$ pairing gaps of Cao et al.~\cite{cao06} (in MeV) for (a) pure 
neutron matter and (b) charge-symmetric nuclear matter, shown as a function of
the appropriate Fermi wave number (in fm$^{-1}$) (see text). 
The results of their 
calculations with self-energy effects included are denoted by filled symbols,
those without by empty symbols. The curves represent our fits to the calculated
points. Figure from Ref.~\cite{gcp16}.}
\label{fig0}
\end{figure}

The pairing force $v^{{\rm pair},q}(\pmb{r_i}, \pmb{r_j})$, as given by 
Eqs.~(\ref{P0}) -- (\ref{P3}), was used directly in the finite-nucleus HFB 
calculations of Ref.~\cite{gcp16} on which the functional BSk31 was based; the 
pairing parameters $f^{\pm}_q, \kappa_q$ and $\varepsilon_\Lambda$ (along with
all the Skyrme parameters of BSk31) are given in Table I of Ref.~\cite{gcp16}.
It is also used in the same way for the BCS calculations of the proton pairing
in the present paper. For neutron pairing, which is calculated in the LDA, 
Eqs.~\eqref{6}-\eqref{8} imply that we need the neutron pairing field  
$\widetilde{\Delta_n}(r)$ corresponding to the complete pairing term (\ref{P1}). 
The LDA amounts to locally solving the BCS gap equations for homogeneous 
nuclear matter with neutron density $\widetilde{n_n}(r)$ and proton density 
$\widetilde{n_p}(r)$ using the complete pairing interaction with gradient term
setting $f_n^\pm=1$. 
This can be achieved by simply inverting Eq.~\eqref{P2} 
using the full pairing interaction~\eqref{P1} instead of $v^{\mathrm{hom}, q}$, and this 
leads to 
\beqy\label{P9}
\widetilde{\Delta_n}(r) = 2\epsilon_{Fn}(r)\exp\left[\frac{2\pi^2\hbar^2}{v^{{\rm pair},n}_{\Lambda}(r)M_n^*[\widetilde{n_n}(r),\widetilde{n_p}(r)]k_{Fn}(r)}
\right] \quad ,
\eeqy
in which we have introduced the effective pairing strength 
\beqy\label{P10}
v^{{\rm pair},q}_{\Lambda}(r)=\left[\frac{1}{v^{{\rm pair},q}(r)} + \frac{M_q^*[\widetilde{n_n}(r),\widetilde{n_p}(r)] k_{Fq}(r)}{4\pi^2 \hbar^2}\Lambda\left(\frac{\varepsilon_\Lambda}{\epsilon_{Fq}(r)} \right) \right]^{-1} \quad .
\eeqy

With the BSk24 functional adopted in our previous works,  there was no density-gradient term
in the associated pairing force. If we adopted the same functional here, approximating
the effective mass by the bare mass in the local Fermi energy entering Eq.~\eqref{P3} as discussed
above, the neutron pairing field $\widetilde{\Delta_n}(r)$ would reduce to the local pairing gaps
$\Delta_n[\widetilde{n_n}(r),\widetilde{n_p}(r)]$, as given by Eq.~\eqref{P6}, and we could
substitute them directly into Eq.~\eqref{6}, without ever constructing any actual pairing force, 
as implemented in Refs.~\cite{sp20,sp21}.

This short-cut is no longer possible now, and having determined the first term
on the right-hand side of Eq.~(\ref{P1}) from the {\it ab initio} gaps of
Cao et al.~\cite{cao06} we now have to add to it the gradient term to construct
the complete pairing force (\ref{P1}) entering Eq.~\eqref{P10}. 
The corresponding neutron pairing field $\widetilde{\Delta_n}(r)$ obtained from 
Eq.~\eqref{P9} no longer coincides with the local pairing gaps 
$\Delta_n[\widetilde{n_n}(r),\widetilde{n_p}(r)]$. 

Cao et al.~\cite{cao06} only calculated the gaps $\Delta_q(n_n, n_p)$ for
pure neutron matter (NM) and charge-symmetric nuclear matter (SM); their
results are shown in Fig.~\ref{fig0}. For arbitrary asymmetry $\eta \equiv
(n_n - n_p)/(n_n + n_p)$ we interpolated~\cite{gcp09} according to
\beqy \label{P6}
\Delta_q(n_n, n_p) = \Delta_\mathrm{SM}(n)(1 - |\eta|) \pm
\Delta_\mathrm{NM}(n_q)\eta\frac{n_q}{n}     \quad  ,
\eeqy
where we take the upper (lower) sign for $q = n (p)$.
Note that this interpolation satisfies the charge symmetry of the nuclear interactions
so that the proton  pairing gap in pure proton matter at density $n$ is the same as the
neutron pairing gap in neutron matter at the same  density $n$, i.e.  $\Delta_p(0, n)=
\Delta_n(n,0)=\Delta_{NM}(n)$.
We parametrize the curves $\Delta_{SM}(n)$ and $\Delta_{NM}(n)$ of
Fig.~\ref{fig0}, thus
\bmlet
\beqy \label{P7a}
\Delta_\mathrm{NM}(n) = \theta(k_m - k_{Fn})\Delta_0\frac{k_{Fn}^2}{k_{Fn}^2 + k_1^2}\,
\frac{(k_{Fn} - k_2)^2}{(k_{Fn} - k_2)^2 + k_3^2}
\eeqy
and
\beqy \label{P7b}
\Delta_\mathrm{SM}(n) = \theta(k_m - k_F)\Delta_0
\frac{k_F^2}{k_F^2 + k_1^2}\,\frac{(k_F - k_2)^2}{(k_F - k_2)^2 + k_3^2}
\quad .
\eeqy
\emlet
Here $k_F = (3\pi^2 n/2)^{1/3}$, $k_{Fn} = (3\pi^2 n_n)^{1/3}$,
$\theta$ is the unit-step Heaviside function and
the parameters $\Delta_0, k_1, k_2, k_3$ and $k_m$ are given in
Table~\ref{tab1}. Note that this parametrization is valid only for the gaps
that Cao et al.~\cite{cao06} calculated with self-energy corrections (filled
symbols in Fig.~\ref{fig0});
it is the gap parametrization that was
adopted in Ref.~\cite{gcp16} (but not shown there) for the construction of
functionals BSk30, BSk31 and BSk32, and is therefore the one adopted here. The
parametrization of the gaps that Cao et al.~\cite{cao06} calculated without
self-energy corrections, and which we adopted for our earlier functionals, in
particular for BSk24, will be found in Ref.~\cite{gcp09}.

\begin{table}
\centering
\caption{Parameters for Eqns. (\ref{P7a}) and (\ref{P7b}).}
\label{tab1}
\begin{tabular} {|c|ccccc|}
\hline
& $\Delta_0$ [MeV] & $k_1$ [fm$^{-1}$] & $k_2$ [fm$^{-1}$] & $k_3$ [fm$^{-1}$] & $k_m$ [fm$^{-1}$] \\
\hline
SM & 11.5586 & 0.489932 & 1.31420 & 0.906146 & 1.31 \\
NM & 3.37968 & 0.556092 & 1.38236 & 0.327517 & 1.38  \\
\hline
\end{tabular}
\end{table}

\section{Results}
\label{results}

Assuming spherical WS cells, we have performed calculations with the new 
functional BSk31 over the inner crust from its interface with the outer
crust, i.e., from the neutron drip point, up to a mean local density $\bar{n}$
of 6.0 $\times 10^{-2}$ fm$^{-3}$, beyond which point proton drip and pasta formation could not have been excluded.

\subsection{Energy and pressure}
\label{eos}

\begin{figure}
\centerline{\includegraphics[width=0.45\columnwidth]{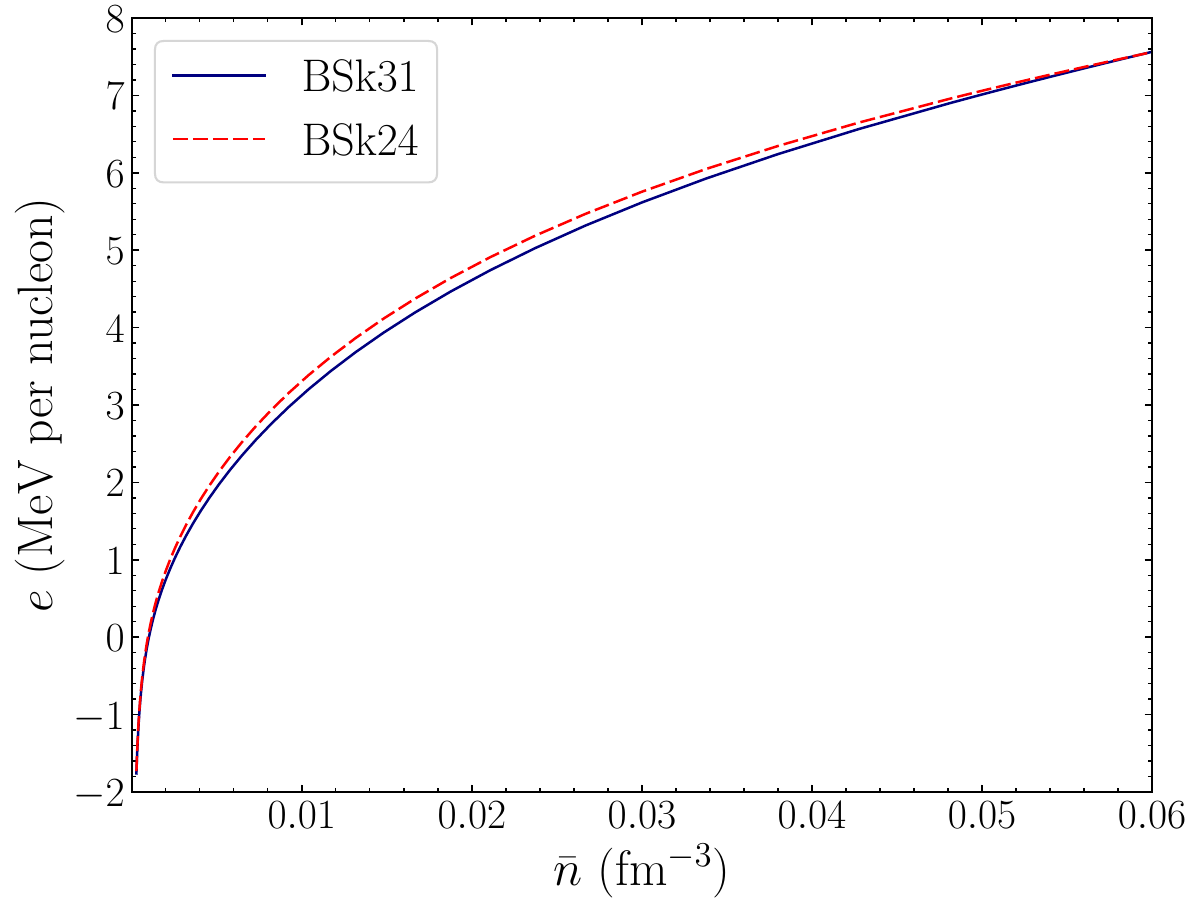}}
\caption{Energy per nucleon (in MeV) in the inner crust of a neutron star as function of mean density $\bar{n}$ (in fm$^{-3}$) for
functionals BSk31 and BSk24.}
\label{fig1}
\end{figure}

In Fig.~\ref{fig1} we show for BSk31 the energy per nucleon $e$ as a function 
of the mean density $\bar{n}$. The same figure also shows results for our previously preferred 
functional, BSk24, calculated with no neutron pairing, as in all our earlier 
calculations with this functional. No significant difference between the
two functionals will be seen, but the fact that the BSk31 rises 
more steeply at high densities suggests that its EoS will be 
stiffer, i.e., the pressure will rise more rapidly with density.
This is confirmed in Fig.~\ref{fig2}. 
Note that the pressure calculated with 
BSk31 now includes the correction due to pairing (see Appendix~\ref{app:pressure}). 

It would be unsafe to assume that these small differences are a result of
neutron pairing being included in BSk31 but not in BSk24, since the two
functionals were fitted to different values of the symmetry coefficient $J$.
Accordingly, to assess the impact of neutron pairing on the EoS we repeat the
calculations of the last two figures for functional BSk31 without neutron 
pairing, denoting it by BSk31(-n). If we replot Figs.~\ref{fig1} and \ref{fig2}
with BSk24 replaced by BSk31(-n) the curves for BSk31 and BSk31(-n) 
would lie very close to each other.
Indeed, as can be seen in Fig.~\ref{fig3}, neutron pairing contributes only a small 
fraction of the total energy per nucleon,  and leads to a gain in energy of 
about 0.1 MeV per nucleon at most.  At all densities the contribution of 
neutron pairing to the pressure 
does not exceed $3\%$, as shown in Fig.~\ref{fig3b}. 
Therefore the changes in the EoS of the inner crust with BSk31 compared with BSk24
are mainly caused by the different symmetry energy.

\begin{figure}
\centerline{\includegraphics[width=0.45\columnwidth]{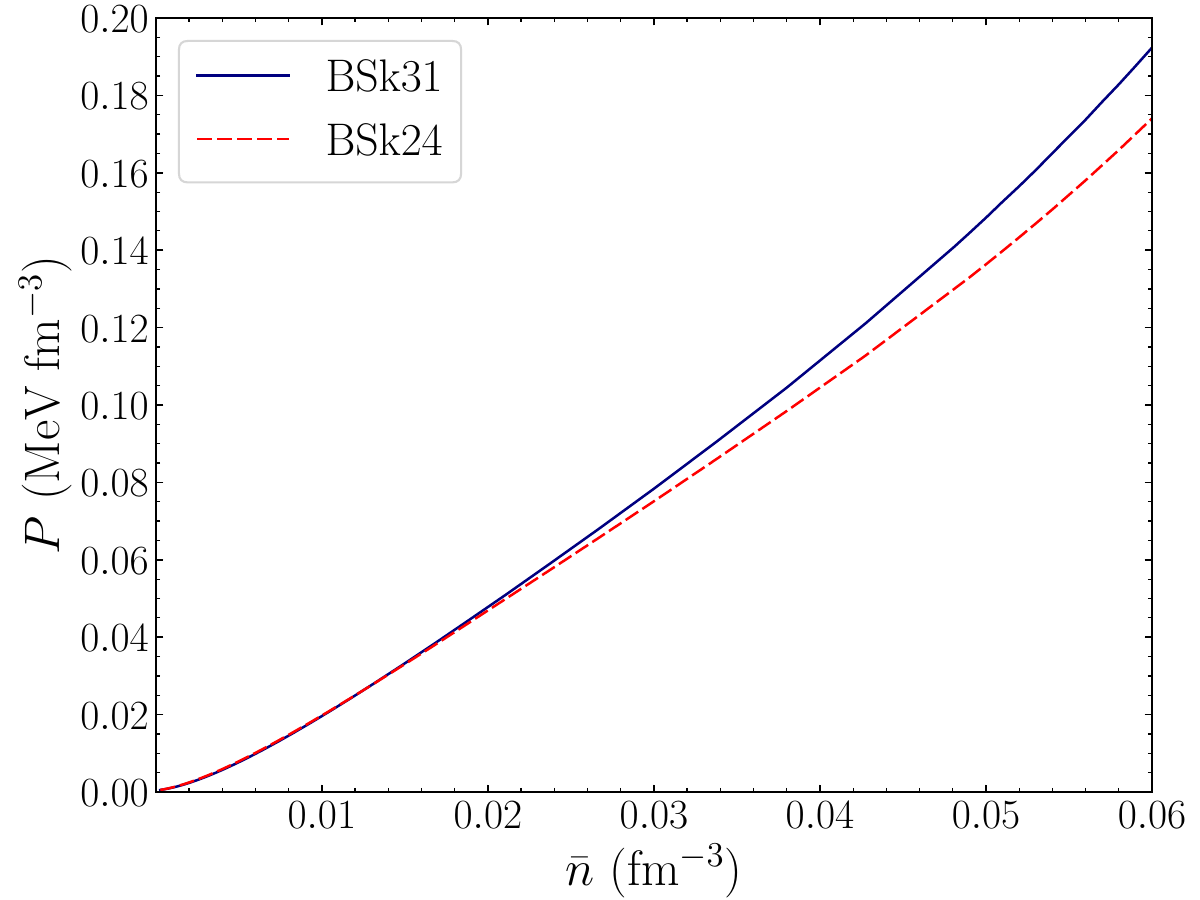}}
\caption{Pressure (in MeV~fm$^{-3}$) in the inner crust of a neutron star as function of mean density $\bar{n}$ (in fm$^{-3}$) for
functionals BSk31 and BSk24.}
\label{fig2}
\end{figure}

\begin{figure}
\centerline{\includegraphics[width=0.45\columnwidth]{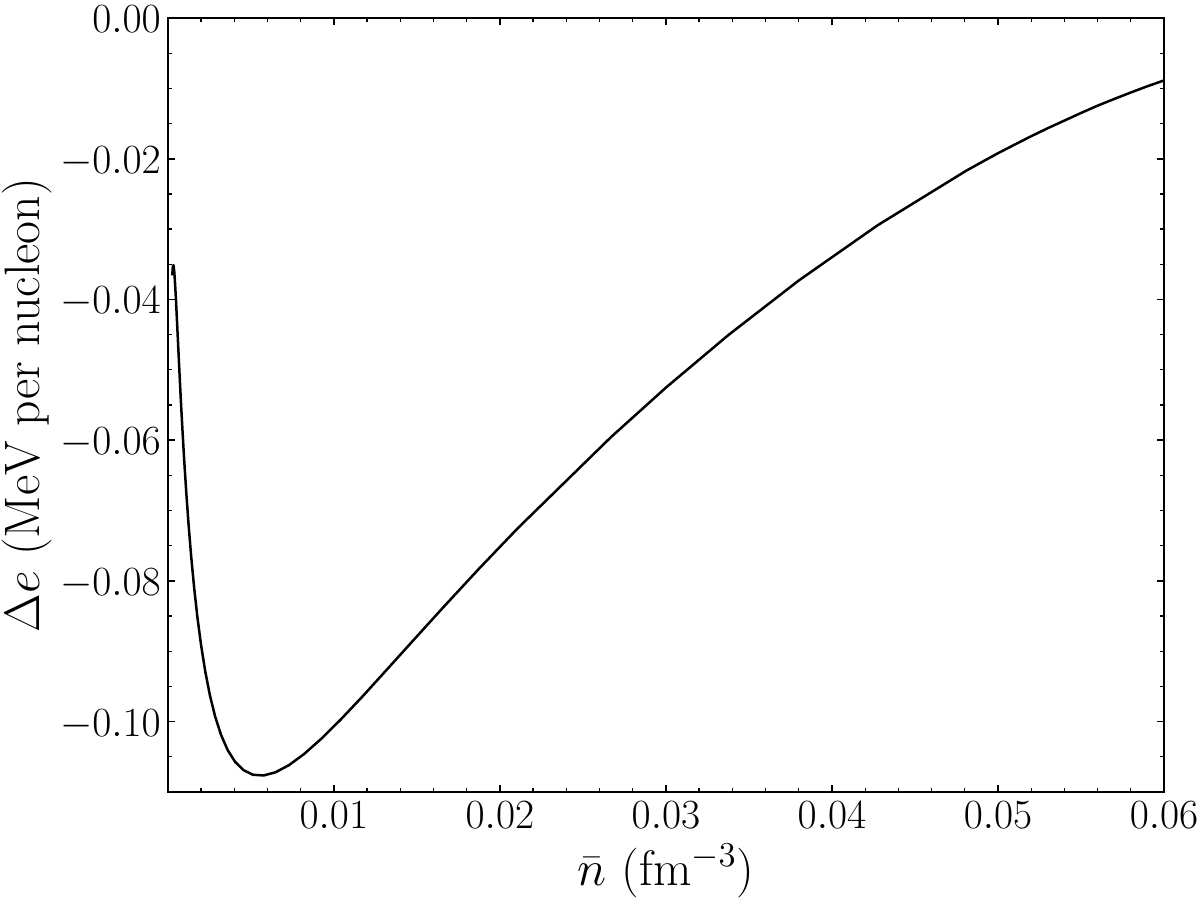}}
\caption{Difference in energy per nucleon (in MeV) for BSk31 calculated with 
and without neutron pairing, as function of mean density $\bar{n}$ 
(in fm$^{-3}$).}
\label{fig3}
\end{figure}

\begin{figure}
\centerline{\includegraphics[width=0.45\columnwidth]{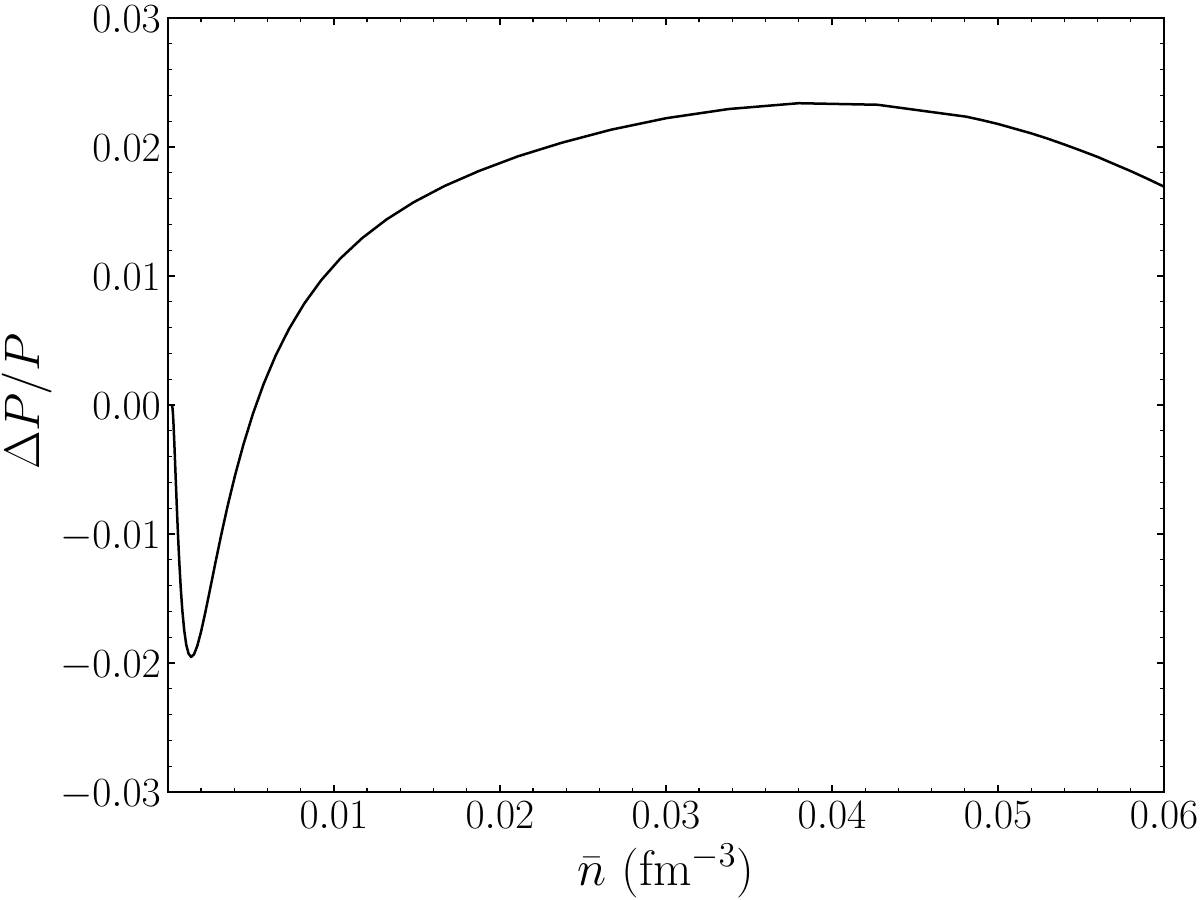}}
\caption{Relative difference in pressure (in MeV~fm$^{-3}$) for BSk31 
calculated with and without neutron pairing as function of mean density 
$\bar{n}$ (in fm$^{-3}$).}
\label{fig3b}
\end{figure}

\subsection{Composition}
\label{compo}

\begin{figure}
\centerline{\includegraphics[width=0.45\columnwidth]{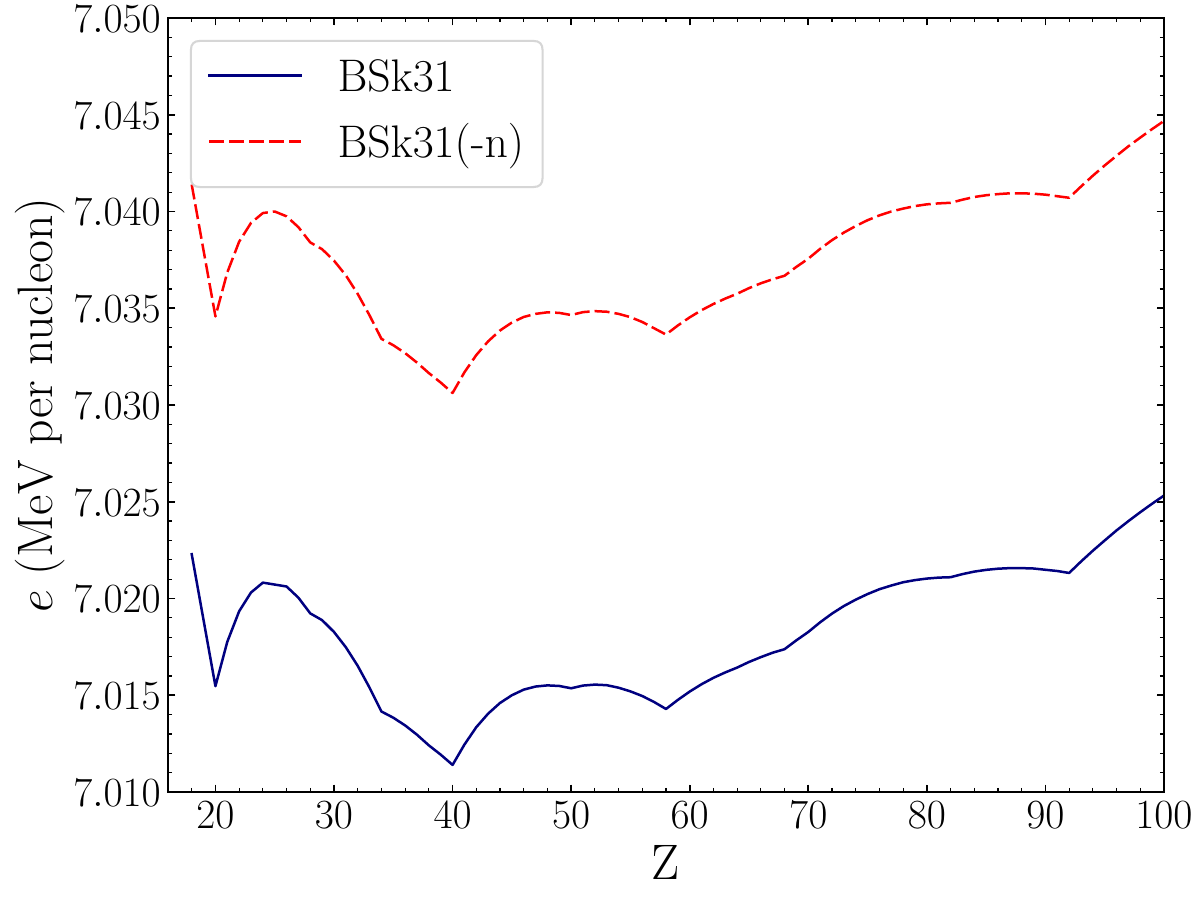}}
\caption{Energy per nucleon (in MeV) as a function of proton number $Z$ at density
$\bar{n}$ of $5.0 \times 10^{-2}$ fm$^{-3}$}
\label{fig4}
\end{figure}

\begin{figure}
\centerline{\includegraphics[width=0.45\columnwidth]{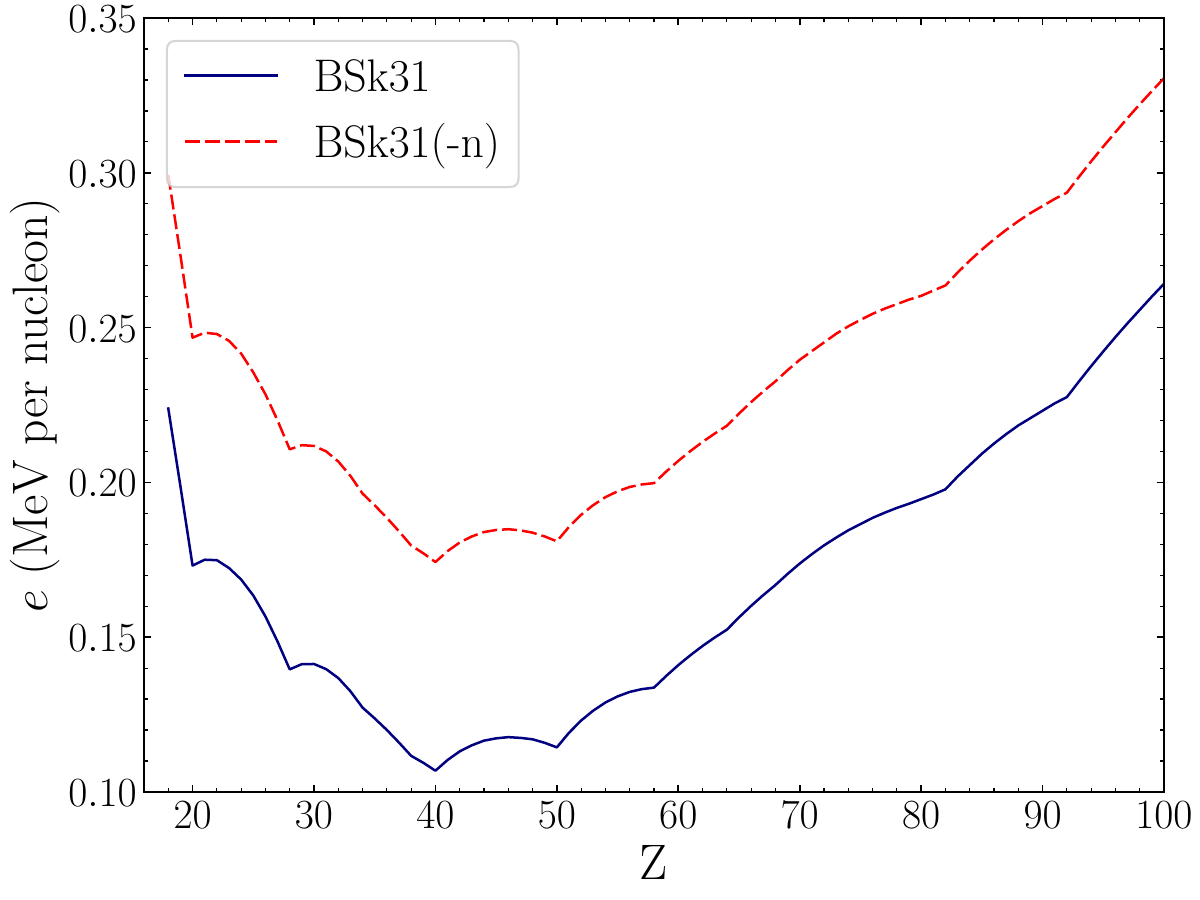}}
\caption{As for Fig.~\ref{fig4} at density
$\bar{n}$ of $1.10967 \times 10^{-3}$ fm$^{-3}$}
\label{fig5}
\end{figure}

\begin{figure}
\centerline{\includegraphics[width=0.45\columnwidth]{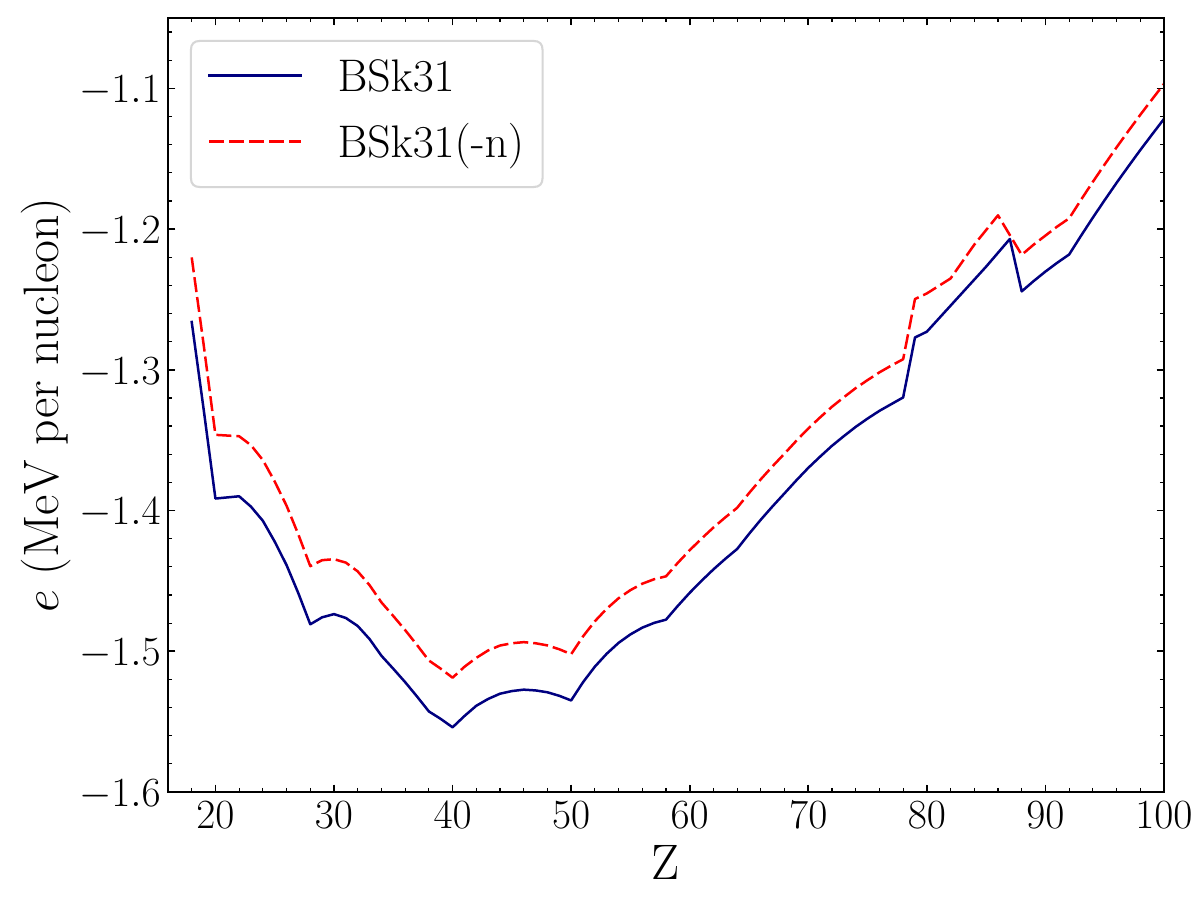}}
\caption{As for Fig.~\ref{fig4} at density
$\bar{n}$ of $3.0375 \times 10^{-4}$ fm$^{-3}$}
\label{fig6}
\end{figure}

In Figs.~\ref{fig4}, \ref{fig5} and \ref{fig6} we show for functionals BSk31 and
BSk31(-n) the variation of the energy per nucleon as a function of proton 
number $Z$ at three different densities spanning the full range considered 
here. We see that the optimal value of the proton number $Z$ is 40 everywhere,
whether or not neutron pairing is
included; the same optimal value of $Z$ for functional BSk24 was found in 
Ref.~\cite{pea18}. However, the energy differences between the different
values of $Z$ are so small that considerable mixtures of different values 
of $Z$ could subsist after the crystallization of the 
crust~\cite{carr2020a,carr2020b}, with pronounced peaks at $Z$ = 40 at all 
densities. Weaker peaks at $Z$ = 20 and 58 will also be found for high 
densities, and at 28 and 50 for lower densities. It is noteworthy that the 
magic numbers 20, 40 and 58 all correspond to the closure of $\ell$-shells 
rather than {\it j}-shells, pointing to the diminished role of spin-orbit 
splitting in the greater homogeneity found at higher densities. 

This insensitivity of the composition of the inner crust to the presence or 
absence of neutron pairing that we find in our ETFSI calculations stands in 
contrast to what was found in the calculations of Refs.~\cite{baldo05,baldo07,grill}. 
The fluctuations in the optimal value of $Z$ found in the HFB calculations
could originate in the spurious shell effects arising from the spherical WS 
cell approximation~\cite{ch07}. The associated errors propagate throughout 
the self-consistent calculations and cannot be completely removed~\cite{grill,pas17}. 

Having established that for functional BSk31 the optimum number $Z$ of protons 
in the spherical WS cell is 40 over the entire density range considered here, whether or
not neutron pairing is included, the question arises as to the optimum number
$N$ of neutrons in the cell, and to what extent this depends on the inclusion 
of neutron pairing. At the same time we note that the optimum value of $N$ will
depend also on the value of the symmetry coefficient $J$ of HNM for the 
functional in question. We thus compare BSk31 not only with BSk31(-n) but also 
with BSk30, for which $J$ = 30 MeV; the latter, like BSk31, will be calculated 
with and without the inclusion of neutron pairing. A complication 
arises from the fact that for BSk30 the optimum value of $Z$ in the inner crust
is equal to 40 only up to $\bar{n} \approx $ 0.002 fm$^{-3}$; for higher 
densities the preferred value of $Z$ is 58, although the energetic advantage
is very slight. We thus study not the optimum value of $N$ but rather that of
the proton fraction $Y_p = Z/(N + Z)$.    

\begin{figure}
\centerline{\includegraphics[width=0.45\columnwidth]{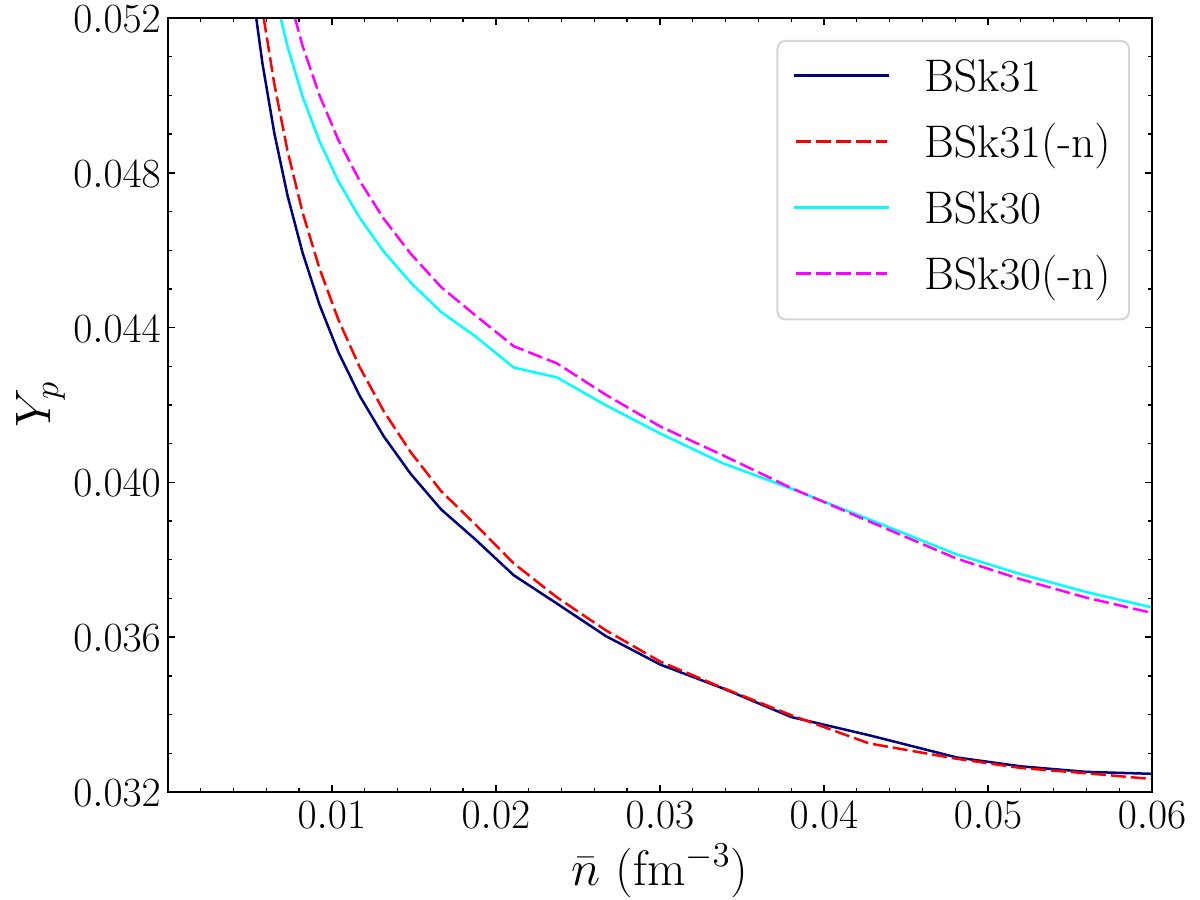}}
\caption{Proton fraction $Y_p$ in the inner crust of a neutron star as function of mean density (in fm$^{-3}$).}
\label{fig7}
\end{figure}

\begin{figure}
\centerline{\includegraphics[width=0.45\columnwidth]{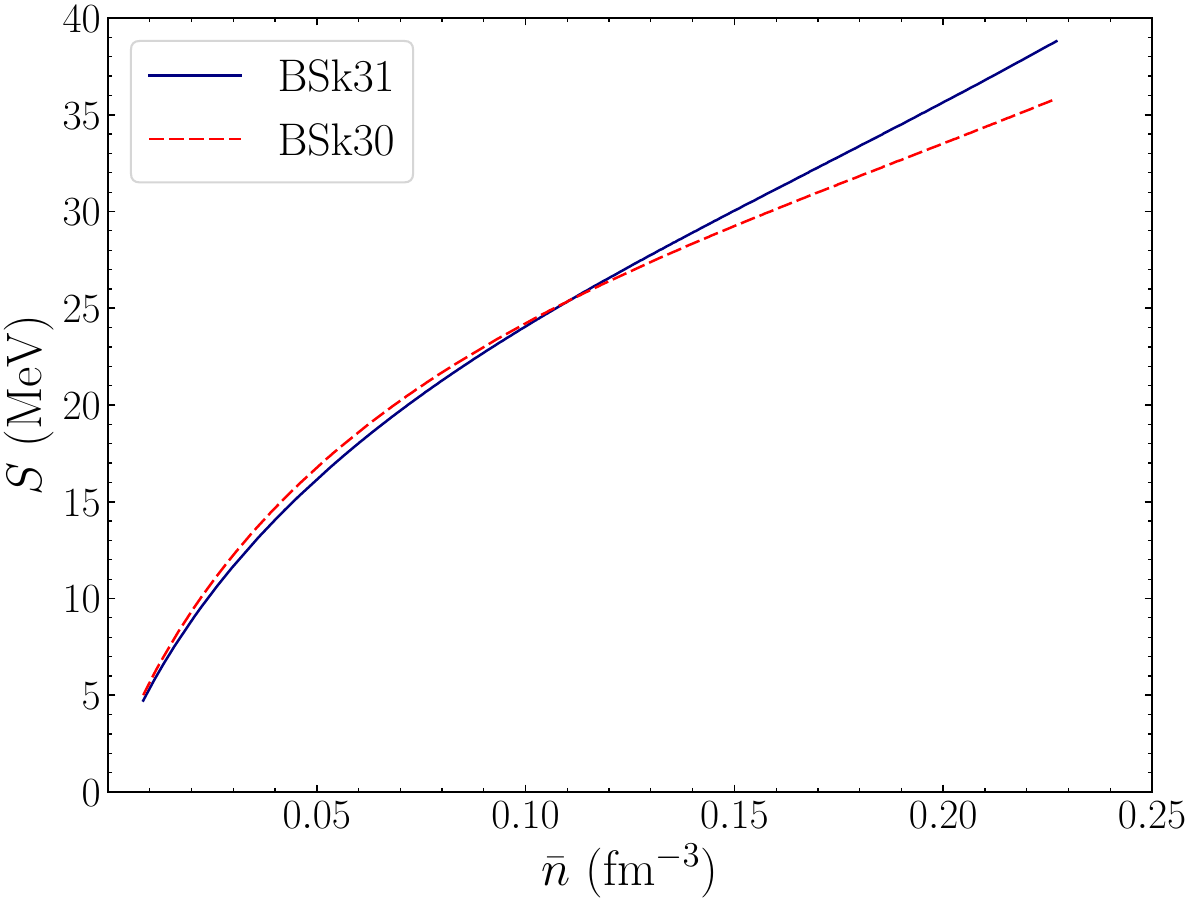}}
\caption{Symmetry energies (in MeV) of BSk31 and BSk30 as a function of density (in fm$^{-3}$).}
\label{fig8}
\end{figure}

The optimum values of $Y_p$ are shown as a function of mean density in 
Fig.~\ref{fig7}. It will be seen that the role of neutron pairing is negligible
compared to that of the symmetry coefficient: the lower value of $J$ is
associated with a higher value of $Y_p$, i.e., fewer neutrons for a given
value of $Z$. This is because at the sub-nuclear densities prevailing in the 
inner crust the symmetry energy obtained with the BSk30-32 functionals 
is {\it greater} for the lower value of $J$, 
as can be seen in Fig.~\ref{fig8}. It would appear that the symmetry 
coefficient $J$ to which the functional in question is fitted has a greater
impact on the number of neutrons in the spherical WS cell than does neutron pairing.
(The kink in both BSk30 curves in the vicinity of $\bar{n}$ = 0.002 fm$^{-3}$
is due to the equilibrium value of $Z$ switching from 40 to 58.) 

\subsection{Neutron pairing field}

\begin{figure}
\centerline{\includegraphics[width=\columnwidth]{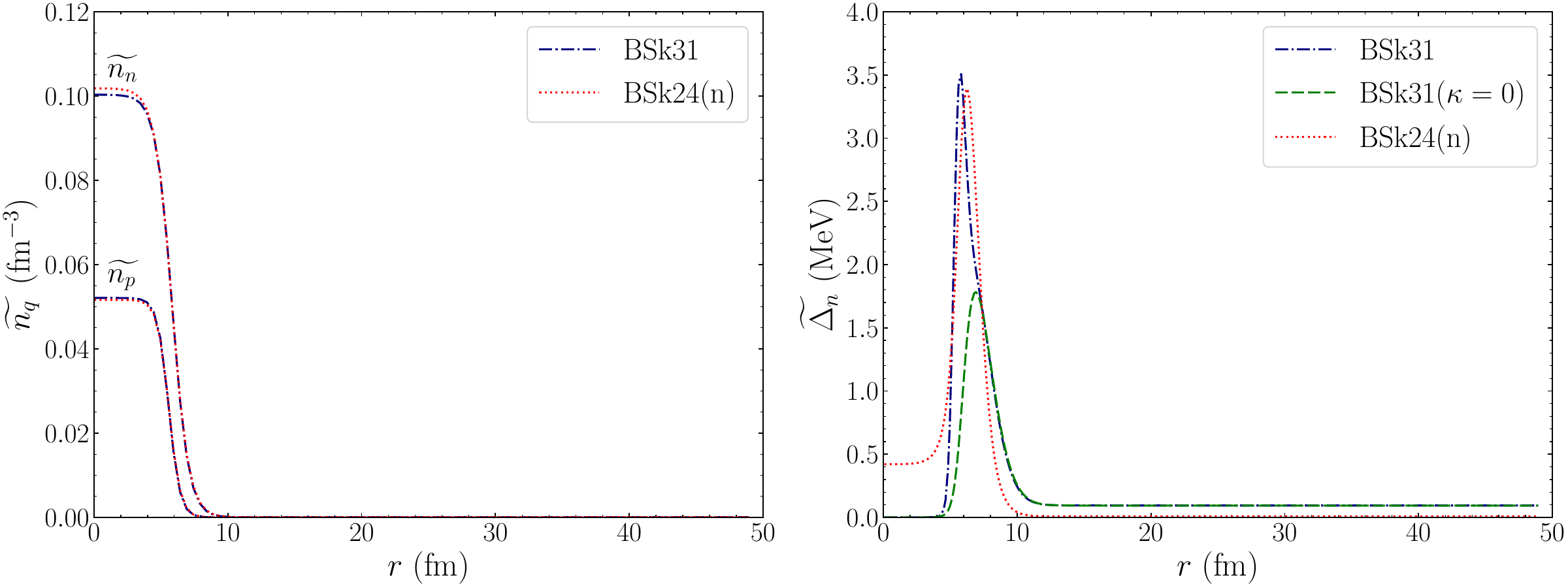}}
\caption{(left panel) Nucleon density distributions in the inner crust of a neutron star at mean density $\bar{n}$ = 3.0 $\times 10^{-4}$ fm$^{-3}$ ($Z = 40$, $A = 148$) as functions of radial position $r$ (in fm) in the spherical WS cell for BSk31 functional (blue dash-dotted lines) and BSk24(n) functional with neutron pairing included (red dotted lines). (Right panel) BSk31 (blue dash-dotted lines) and BSk24(n) (red dotted lines) neutron pairing fields (in MeV) in the spherical WS cell. The pairing field for BSk31 without density gradient term is plotted for comparison (BSk31($\kappa=0$) -- green dashed line). }
\label{fig9}
\end{figure}

\begin{figure}
\centerline{\includegraphics[width=\columnwidth]{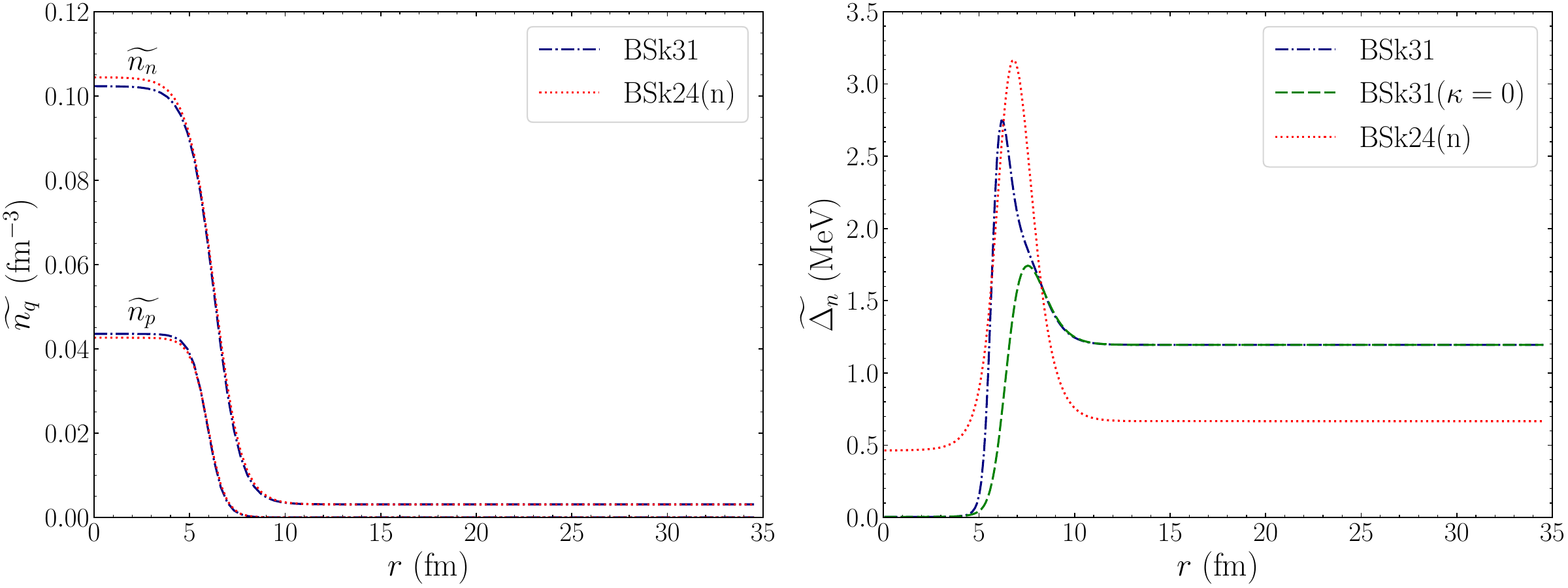}}
\caption{As for Fig.~\ref{fig9} at density $\bar{n}$ = 4.0 $\times 10^{-3}$
fm$^{-3}$ ($Z = 40$, $A = 693$).}
\label{fig10}
\end{figure}

\begin{figure}
\centerline{\includegraphics[width=\columnwidth]{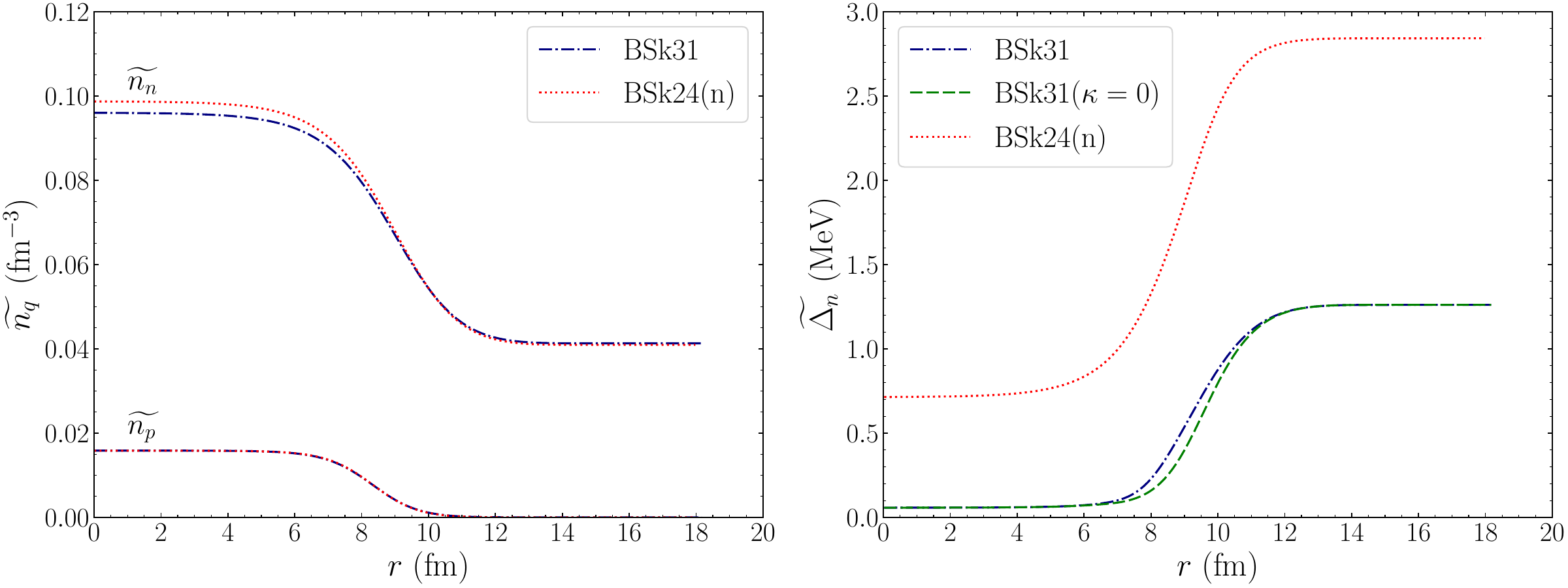}}
\caption{As for Fig.~\ref{fig9} at density $\bar{n}$ = 5.0 $\times 10^{-2}$
fm$^{-3}$ ($Z = 40$, $A = 1222$).}
\label{fig11}
\end{figure}

In Figs.~\ref{fig9}, \ref{fig10} and \ref{fig11} we compare the
nucleon density distributions $\widetilde{n_q}(r)$ and 
neutron pairing fields $\widetilde{\Delta_n}(r)$ generated 
by the functional BSk31 and those generated by BSk24 with its neutron 
pairing taken into account, a feature that we indicate by denoting it as 
BSk24(n). All three figures show the variation of the nucleon density distributions  
and neutron pairing fields with the radial position $r$, Fig.~\ref{fig9} for the low
mean density of $\bar{n}$ = 3.0 $\times 10^{-4}$ fm$^{-3}$, Fig.~\ref{fig10} 
for the intermediate mean density of 4.0 $\times 10^{-3}$ fm$^{-3}$ and 
Fig.~\ref{fig11} for the high mean density of 5.0 $\times 10^{-2}$ fm$^{-3}$. 
In all three  figures, the curves labelled ``BSk31($\kappa$ = 0)" relate to 
calculations performed with functional BSk31 in which the part of the neutron 
pairing term that depends on the density gradient has been removed. To make the
comparisons more meaningful the pairing fields are calculated with the same number of 
neutrons, $N$ = 108 for Fig.~\ref{fig9}, 653 for Fig.~\ref{fig10} and 1182 for 
Fig.~\ref{fig11}, these being optimal values for BSk31; in all cases we 
take $Z$ = 40, this being optimal for both functionals at all densities.  

At the very low mean density $\bar{n}$ of Fig.~\ref{fig9}, close to the 
interface with the outer crust, we see the strong contribution made by the 
density-gradient pairing term in BSk31 in the vicinity of 
$r$ = 6 fm, which is
just the highly inhomogeneous region corresponding to the tail of the cluster 
term of the density distribution~(\ref{2}). 
Remarkably, the maximum value of the neutron pairing field turns out to be 
comparable to that obtained with BSk24(n). To achieve an equally good nuclear 
mass fit, the absence 
of a gradient term in the latter could only be compensated by neglecting 
self-energy effects. 

Outside of this region, both close to the center of the cell and
towards the surface of the cell, the density distribution is relatively
homogeneous and the BSk31 neutron pairing field is determined almost entirely by the 
reference gap of Cao et al.~\cite{cao06},
see Fig.~\ref{fig0}. The BSk24(n) neutron pairing field does not exactly
coincide with the reference gap of Cao et al.~\cite{cao06} because of the substitution of the 
bare mass instead of the effective mass in the pairing force $v^{{\rm pair},q}(\pmb{r})$ (as well as in the 
Fermi energy entering the pairing force) but not in Eqs.~\eqref{P9} and \eqref{P10}.
At the center of the cluster, $k_{F}\approx1.3$\,fm$^{-1}$ and $k_{Fn}\approx1.4$\,fm$^{-1}$ 
correspond to negligibly small SM and NM gaps but only when self-energy effects are included. This explains 
why the neutron pairing field vanishes for BSk31 while it remains non-negligible for BSk24(n). 
At the border of the cell, the matter consists mainly of free neutrons and the left panel of Fig.~\ref{fig0} is 
highly relevant. There, $k_{Fn}\approx0.1$\,fm$^{-1}$ and for this wave number the pairing gaps 
$\Delta_\mathrm{NM}$ are almost equal for BSk31 and BSk24(n), as can be seen in the left panel of Fig.~\ref{fig0}.
The differences in the neutron pairing field come from the different treatments of the effective mass 
in the pairing force.   

The situation depicted at the intermediate value of $\bar{n}$ in 
Fig.~\ref{fig10} is qualitatively unchanged, but at the much higher mean 
densities $\bar{n}$ close to the mantle the spherical WS cell is everywhere 
more 
homogeneous and the density-gradient term in BSk31 is now almost ineffective, 
as can be seen in Fig.~\ref{fig11} comparing the blue dash-dotted and green 
dashed curves. Thus the calculated neutron pairing field everywhere strongly 
resembles the HNM reference gaps. In particular, the strong increase of the neutron 
pairing field outside clusters stems
from the fact that the value of $k_{Fn}\approx1.0$\,fm$^{-1}$ corresponds to the 
peak of $\Delta_\mathrm{NM}$ shown in the left panel of Fig.~\ref{fig0}.  

It is apparent from Figs.~\ref{fig9}, \ref{fig10} and \ref{fig11} that it is 
only at higher mean densities $\bar{n}$ that the BSk24(n) neutron pairing, which 
is based on the HNM pairing calculated without self-energy corrections, 
can be said to be stronger than the BSk31 neutron pairing, but these of course 
are the densities relevant to the nuclear mass fits.

It is worth noting that Figs.~\ref{fig9}, \ref{fig10} and \ref{fig11} show that
for all values of the mean density $\bar{n}$ the neutron pairing fields remain constant 
for some distance below the cell surface. This is because the cluster 
term in ~(\ref{2}) becomes vanishingly small for relatively small values of $r$,
beyond which the pairing fields are determined by the background parameters 
$n_{\mathrm{B}n}$ and $n_{\mathrm{B}p}$.

\section{Conclusion}
\label{concl}

In this paper, we have further improved our ETFSI treatment of the inner crust 
of a neutron star by adding neutron pairing within the LDA, using the 
functional BSk31~\cite{gcp16}. Since the pairing term of this functional 
depends on the density gradient a fairly extensive modification of our previous
treatment is necessary. 

Pending the completion of our improved treatment of pasta, our present 
calculations are confined to densities below 0.06 fm$^{-3}$ where clusters 
are expected to be quasispherical. We have found that neutron pairing has a 
marginal impact on the composition.  The equilibrium proton number $Z$ remains 
equal to 40, and the number of neutrons in the spherical WS cell is only slightly 
altered, a conclusion that is at variance with results obtained from HFB
and HF+BCS calculations~\cite{baldo05,baldo07,grill}.  The conclusions of those calculations, 
however, are rendered questionable by spurious shell effects arising from the 
spherical WS cell approximation~\cite{ch07}.  
Comparing results obtained with BSk30 and BSk31 from the same family  but 
fitted to different values of $J$ shows that the symmetry energy plays a more 
important role for the composition than does pairing.  As expected, the neutron 
pairing correction to the energy per nucleon is small, being of order $0.1$ MeV at most. 
Likewise, the correction to the pressure does not exceed $3\%$.

We have shown that while the functional BSk31 is equivalent in many 
respects to our previously preferred functional BSk24~\cite{gcp13}, it has, as expected,
quite different neutron-pairing properties. 
Being more realistically based by including 
not only medium polarization but also self-energy effects,
BSk31 is more suitable for the study of neutron superfluidity in the inner 
crust of neutron stars.
In particular, our calculations suggest that the neutron superfluid dynamics
could be qualitatively different, which could have important implications
for the global dynamics of neutron stars and the interpretation of
astrophysical phenomena, such as pulsar frequency glitches~\cite{zho22}. The
key point is the conclusion that while with the functional BSk24(n) the neutron
pairing field penetrates the clusters everywhere in the inner crust, with BSk31
the pairing field vanishes inside clusters in all regions of the inner crust
but the deepest (see Figs.~\ref{fig10}-\ref{fig11}). This suggests
that the neutron superfluid can flow through the clusters for the functional BSk24(n), but
not for the functional BSk31, except close to the interface with the core of
the star. To analyze the implication that the neutron superfluid must flow
around the clusters in the case of the functional BSk31 one must go beyond the
ETFSI+pairing approach adopted here and work within the framework of the
time-dependent HFB approach. However, a considerable economy in the computer
time consumed by such calculations could be achieved by using the WS-cell
parameters (composition and nucleonic distributions) determined in the
calculations of the present paper.

Since proton superconductivity only occurs at densities higher
than those considered here a comparable study of the proton-pairing properties and proton superconductivity 
of the functional BSk31 is postponed to a later paper.    

\begin{acknowledgments}
The work of N.N.S. was 
financially supported by the FWO (Belgium) and the Fonds de la Recherche Scientifique (Belgium) under the Excellence of Science (EOS) programme (project No. 40007501). This work also received funding from the Fonds de la Recherche Scientifique (Belgium) under Grant No. PDR T.004320 and IISN 4.4502.19. 
\end{acknowledgments}

\appendix
\section{Generalized Strutinsky-integral theorem}
\label{app:Strutinsky}

The aim of this appendix is to provide for the first time a rigorous derivation
for the SI correction 
with BCS pairing~\eqref{4} originally introduced in Ref.~\cite{pear91} in the context 
of finite nuclei using heuristic arguments. 
To this end, we generalize the proof of the SI theorem given in 
Appendix C of Ref.~\cite{ons08} within the pure Hartree-Fock (HF) approach, without 
pairing.

In the HFB method (see, e.g., Ref.~\cite{rs80}), the 
energy $E_\textrm{HFB}$ of the (normalized) 
ground-state $\vert \Psi\rangle$ is expressed as a function of the so-called 
normal and abnormal density matrices, defined by (we drop here the nucleon 
label $q$) 
\bmlet
\beqy\label{A1a}
n_{ij} = <\Psi|c_j^\dagger c_i|\Psi> = \sum_k V_{ik}^*V_{jk} = n_{ji}^*
\eeqy
and
\beqy\label{A1b}
\kappa_{ij} = <\Psi|c_jc_i|\Psi> = \sum_k V_{ik}^*U_{jk} = - \kappa_{ji}  \quad  ,
\eeqy
\emlet 
respectively, with $c^\dagger_i (c_j)$ denoting creation (destruction) operators for nucleons 
in such states. Here we are working in a fixed basis of discrete s.p. states labelled by $i$, $j$ , etc., e.g., 
an oscillator basis (the isospin charge type is implicit in the label). 
The matrices $U_{ij}$ and $V_{ij}$ can be obtained from the HFB equations 
\beqy\label{A2}
\sum_j \begin{pmatrix} h_{ij}-\lambda \delta_{ij} & \Delta_{ij} \\ -\Delta_{ij}^* & -h^{*}_{ij} 
+ \lambda \delta_{ij} \end{pmatrix}\begin{pmatrix} U_{jk} \\ V_{jk} \end{pmatrix} =
E_k \begin{pmatrix} U_{ik} \\ V_{ik} \end{pmatrix}  \quad ,
\eeqy
where $E_k$ are the quasi-particle energies, $\lambda$ is the chemical potential, 
$h_{ij}$ are the matrix elements of the self-consistent s.p. Hamiltonian
\bmlet
\beqy\label{A3a}
h_{ij} = \frac{\partial\,E_{\rm HFB}}{\partial n_{ji}} = 
h_{ji}^{*} \, , 
\eeqy
and $\Delta_{ij}$ are the matrix elements of the pairing potential
\beqy\label{A3b}
\Delta_{ij} = \frac{\partial\,E_{\rm HFB}}{\partial\kappa^*_{ij}} =
-\Delta_{ji} \, .
\eeqy
\emlet
For zero-range density-dependent effective interactions, such as those 
considered here, the HFB energy is given by 
\beqy\label{A4}
E_{\rm HFB} = {\rm Tr}\left(t n +\frac{1}{2}\Gamma n-
\frac{1}{2}\Delta\kappa^*\right) \quad ,
\eeqy
in which Tr denotes the trace, $t_{ij}$ are the matrix elements of the kinetic-energy operator 
$-\hbar^2\pmb{\nabla}^2 / 2 M$ ($M$ denoting the nucleon mass), while 
\bmlet
\beqy\label{A5a}
\Gamma_{kl} = \sum_{ij}\bar{v}_{ki,lj}^{\rm Sky}\,n_{ji} + \sum_{ij}\bar{v}_{ki,lj}^{\rm Coul}\,n_{ji} \, ,
\eeqy
and
\beqy\label{A5b}
\Delta_{kl} = \frac{1}{2}\sum_{ij}\bar{v}_{kl,ij}^{\rm pair}\,\kappa_{ij} 
\quad ,
\eeqy
\emlet 
where $\bar{v}_{ki,lj}^{\rm Sky}$, $\bar{v}_{ki,lj}^{\rm Coul}$ and $\bar{v}_{ki,lj}^{\rm pair}$ are the 
antisymmetrized matrix elements of the Skyrme, Coulomb and pairing interactions respectively. 
Using Eq.~(\ref{A3a}), the matrix elements of the self-consistent s.p. Hamiltonian are thus given by
\beqy\label{A6}
h_{ij} = t_{ij} + \Gamma_{ij}+ h_{ij}^{\rm rear} \, , 
\eeqy
where we have introduced matrix elements of the rearrangement s.p. field
\beqy\label{A7}
h_{ij}^{\rm rear} \equiv \frac{1}{2} \sum_{klpm}\left(
 \frac{\partial \bar{v}^{\rm Sky}_{kl,pm}} {\partial n_{ji}}n_{ml}n_{pk} - 
\frac{1}{2} \frac{\partial\bar{v}_{kl,pm}^{\rm pair}}{\partial n_{ji}}
\kappa_{pm}\kappa^*_{lk}\right)   \quad .
\eeqy
For the relation between this matrix formulation of the HFB equations and the coordinate space formulation, see, e.g., 
Appendix A of Ref.~\cite{cgp08}. 

Multiplying the second row of Eq.~(\ref{A2}) by $V_{ik}^*$ using 
Eqs.~(\ref{A1a}), (\ref{A1b}) and (\ref{A6}), and summing over $k$ and $i$ 
yields
\beqy\label{A8}
{\rm Tr}\left(\Delta^* \kappa - t n-\Gamma n-h^{\rm rear}n\right)=\sum_{ki} (E_k-\lambda)| V_{ik}|^2\, .
\eeqy
Inserting this expression into Eq.~(\ref{A4}) leads to 
\beqy\label{A9}
E_{\rm HFB}=\sum_{ki} (\lambda- E_k)| V_{ik}|^2 - E_\textrm{pair}-\frac{1}{2}{\rm Tr}\left(\Gamma n \right)-{\rm Tr}\left(h^{\rm rear}n\right)  \quad ,
\eeqy
with the pairing energy 
\beqy\label{A10}
E_{\rm pair}=-(1/2){\rm Tr}\left(\Delta^* \kappa\right)\leq 0 \quad . 
\eeqy

In the BCS approximation, the matrix elements of the pairing potential are supposed to take the form 
\beqy\label{A11}
\Delta_{kl} =\delta_{\bar k l}\, \Delta_{k\bar k}  \quad ,
\eeqy
$\bar k$ denoting the time-reversed of the state $k$ 
in the basis for which the s.p. Hamiltonian is diagonal,
\beqy\label{A12}
h_{ij}=\epsilon_i\, \delta_{ij} \quad .
\eeqy
Solving now the HFB equations~(\ref{A2}) with this ansatz and 
making use of the anticommutation and unitarity relations between the $U$ and $V$ matrices 
(see, for example, Eqs. (7.5) of Ref~\cite{rs80}), we find
\bmlet
\beqy
\label{A13a}
E_k = \sqrt{(\epsilon_k-\lambda)^2+\Delta_k^2} \quad ,
\eeqy
\beqy\label{A13b}
U_{kk}=U_{\bar k\bar k} = \frac{1}{\sqrt{2}}\left(1+\frac{\epsilon_k-\lambda}{E_k}\right)^{1/2} \quad ,
\eeqy
and
\beqy\label{A13c}
V_{k\bar k}=-V_{\bar k k} = \frac{1}{\sqrt{2}}\textrm{sign}(\Delta_{\bar k k})\left(1-\frac{\epsilon_k-\lambda}{E_k}\right)^{1/2}\equiv V_k \quad ,
\eeqy
\emlet
where we have introduced the pairing gaps $\Delta_k\equiv |\Delta_{\bar k k}|$
and we have adopted the usual phase convention. Substituting Eqs.~(\ref{A13b}) 
and (\ref{A13c}) into Eqs.~(\ref{A1a}) and (\ref{A1b}) leads to the familiar 
expressions for the normal and abnormal density matrices,
\beqy\label{A14}
n_{kl}= (V_{k \bar k})^2\, \delta_{kl}\, , \hskip 0.5cm 
 \kappa_{kl}=V_{k \bar k} U_{k k}\, \delta_{\bar k l} \, .
\eeqy  
Using Eqs.~(\ref{A14}) and (\ref{A5b}) leads to the BCS gap equations
\beqy\label{A15}
\Delta_k = -\frac{1}{4} \sum_{l}\bar{v}_{k \bar k,l \bar l}^{\rm pair} \frac{\Delta_{l}}{E_{l} }\, .
\eeqy
The pairing energy, as given by Eq.~(\ref{A10}), reduces to Eq.~\eqref{5}. 

In the absence of pairing (HF limit), 
$\kappa_{ij}=0$, and therefore $\Delta_{ij}=0$ in any basis, in particular 
in the basis of the s.p. Hamiltonian where $h_{ij}=\breve{\epsilon}_i\, \delta_{ij}$. Note that 
$\breve{\epsilon}_i$ does not generally coincide with $\epsilon_i$ because the effective pairing force contributes 
to the s.p. mean field via rearrangement terms, see Eq.~(\ref{A7}). In the following, we will denote a quantity $Q$ obtained in the 
HF approximation by $\breve{Q}$. Using Eq.~(\ref{A13c}) yields 
$\breve{V}_{i}^2=1$
if $\breve{\epsilon}_i\leq \breve{\lambda}$ and $0$ otherwise. 
Likewise, Eq.~(\ref{A13a}) leads to $\breve{E}_i=|\breve{\epsilon}_i-\breve{\lambda}|$. Using Eq.~(\ref{A9}), we find for the HF
ground-state energy
\beqy\label{A17}
E_{\rm HF}=\sum_{i} n_i \breve{\varepsilon}_i -\frac{1}{2}{\rm Tr}\left(\breve{\Gamma}\breve{n}\right)-{\rm Tr}\left(\breve{h}^{\rm rear}\breve{n}\right) \quad ,
\eeqy
where $n_i\equiv \breve{V}_i^2$ is the Fermi-Dirac distribution. We shall neglect the (small) differences between the HF and HF+BCS
density matrices, i.e. $\breve{n}_{ij}\approx n_{ij}$ (hence also $\breve{\Gamma}_{ij}\approx\Gamma_{ij}$), as well as the differences in the 
rearrangement s.p. field, i.e. $\breve{h}_{ij}^{\rm rear} \approx h_{ij}^{\rm rear}$ therefore $\breve{\epsilon}_i\approx \epsilon_i$. 
With $\breve{\lambda}\approx\lambda$, the HF+BCS ground state energy can finally be expressed as 
\beqy\label{A18}
E_{\rm HF+BCS}\approx E_{\rm HF}-E_{\rm pair}+ \sum_{i} V_{i}^2 (\lambda-E_i) - \sum_{i} n_i \epsilon_i \quad .
\eeqy

On the other hand, the HF energy can be obtained from the 
Strutinsky-integral theorem (see, e.g., Appendix C of Ref.~\cite{ons08}) 
\beqy\label{A19}
E_{\rm HF} \approx E_{\rm ETF}  + \delta E_{\rm HF} \, ,
\eeqy
where $E_{\rm ETF}$ is the ``macroscopic'' energy, as calculated using the ETF method~\cite{bbd, opp97}, 
and $\delta E_{\rm HF}$, given by 
\beqy\label{A20}
\delta E_{\rm HF} = \sum_{k} n_k \widetilde{\epsilon}_{k} -\int d^3\pmb{r}\biggl(\frac{\hbar^2}{2\widetilde{M}^*}\widetilde{\tau} +
\widetilde{n}\widetilde{U}+\widetilde{\pmb{J}}\cdot\widetilde{\pmb{W}}\biggr)  \, ,
\eeqy 
accounts for shell corrections. 
Inserting Eq.~(\ref{A19}) in Eq.~(\ref{A18}) using Eq.~\eqref{A20}, we find 
\beqy\label{A21}
E_{\rm HF+BCS}\approx E_{\rm ETF}+\delta E_{\rm HF+BCS} \quad ,
\eeqy
with 
\beqy\label{A22}
\delta E_{\rm HF+BCS}=\sum_{k} V_{k}^2 (\lambda-E_k) - \int d^3\pmb{r}\biggl(\frac{\hbar^2}{2\widetilde{M}^*}\widetilde{\tau} +
\widetilde{n}\widetilde{U}+\widetilde{\pmb{J}}\cdot\widetilde{\pmb{W}}\biggr)+\sum_k \frac{\Delta_k^2}{4 E_k} \quad .
\eeqy
Using Eqs.~(\ref{A13a}) and (\ref{A13c}), this expression can be written in the equivalent form 
\beqy\label{A23}
\delta E_{\rm HF+BCS}=\sum_{k} V_{k}^2 \epsilon_k - \int d^3\pmb{r}\biggl(\frac{\hbar^2}{2\widetilde{M}^*}\widetilde{\tau} +
\widetilde{n}\widetilde{U}+\widetilde{\pmb{J}}\cdot\widetilde{\pmb{W}}\biggr) - \sum_k \frac{\Delta_k^2}{4 E_k} \quad ,
\eeqy
which is the correction given by Eqs.~(\ref{4}) and (\ref{5}).

\section{Pairing correction to the pressure}
\label{app:pressure}

As shown in Ref.~\cite{pcgd12}, the pressure of any crustal layer is the same as that obtained in an homogeneous mixture of 
neutrons, protons, and electrons with densities $n_{\mathrm{B}n}$, $n_{\mathrm{B}p}$, and $n_e$ respectively. For the inner crust region considered 
here, protons remain bound inside clusters therefore $n_{\mathrm{B}p}\approx 0$. Therefore, the proton pairing correction to the pressure 
given by Eq.(B25) of Ref.~\cite{pcgd12} can be safely neglected. The neutron pairing correction can be calculated from the condensation 
energy density of pure neutron matter at density $n_{\mathrm{B}n}$, as 
\beqy 
\delta P = n_{\mathrm{B}n}^2 \frac{d(\mathcal{E}_{cond,n}/n_{\mathrm{B}n})}{d n_{\mathrm{B}n}}\, , 
\eeqy 
where 
\beqy 
\mathcal{E}_{cond,n}=-\frac{3}{8} n_{\mathrm{B}n} \frac{\Delta_{NM}(n_{\mathrm{B}n})^2}{\epsilon_{Fn}}\, ,
\eeqy 
\beqy 
\epsilon_{Fn}=\frac{\hbar^2}{2 M_n^*(n_{\mathrm{B}n})}\left(3\pi^2 n_{\mathrm{B}n}\right)^{2/3} \, . 
\eeqy 
The pressure correction for the generalized Skyrme functionals adopted in this work~\cite{gcp16} can be expressed as 
\beqy 
\delta P = \frac{n_{\mathrm{B}n}^{1/3}\Delta_{NM}(n_{\mathrm{B}n}) \left[3 n_{\mathrm{B}n} \Delta_{NM}(n_{\mathrm{B}n}) B_n'(n_{\mathrm{B}n}) + 2 B_n(n_{\mathrm{B}n}) (\Delta_{NM}(n_{\mathrm{B}n}) - 3 n_{\mathrm{B}n} \Delta_{NM}'(n_{\mathrm{B}n}))\right]}{8(3 \pi^2)^{2/3} B_n(n_{\mathrm{B}n})^2} \, , 
 \eeqy 
where we have introduced 
\beqy 
B_n(n_{\mathrm{B}n})\equiv\frac{\hbar^2}{2 M_n^*(n_{\mathrm{B}n})}=\frac{\hbar^2}{2 M_n}+\frac{1}{8} \left[t_1(1 - x_1) + 3 t_2 (1 + x_2) + t_4 (1 - x_4) n_{\mathrm{B}n}^\beta +  3 t_5 (1 + x_5) n_{\mathrm{B}n}^\gamma\right] n_{\mathrm{B}n} \, . 
\eeqy 
The derivative of $B_n(n_{\mathrm{B}n})$ and $\Delta_{NM}(n_{\mathrm{B}n})$ with respect to $n_{\mathrm{B}n}$, indicated by a prime, can be easily calculated 
analytically from the expression above and Eq.~\eqref{P7a} (with $k_{Fn}=(3\pi^2 n_{\mathrm{B}n})^{1/3}$) respectively, and are not explicitly given.


\begin{thebibliography}{99}

\bibitem{bc18} D.~Blaschke and N.~Chamel, in \textit{The Physics and
Astrophysics of Neutron Stars}, edited by L.~Rezzolla, P.~Pizzochero, D.~Jones,
N.~Rea, and I.~Vida\~{n}a, Astrophysics and Space Science Library Vol. 457
(Springer, Berlin, 2018), pp.~337--400.

\bibitem{pea18}J.~M.~Pearson, N.~Chamel, A.~Y.~Potekhin, A.~F.~Fantina,
C.~Ducoin, A.~K.~Dutta, and S.~Goriely, MNRAS, \textbf{481}, 2944 (2018)
[Erratum: MNRAS, \textbf{486}, 768 (2019)].

\bibitem{gcp13}S.~Goriely, N.~Chamel, and J.~M.~Pearson,
Phys.\ Rev. C \textbf{88}, 024308  (2013).

\bibitem{ame12} G.~Audi, M.~Wang, A.~H.~Wapstra, F.~G.~Kondev, M.~MacCormick,
X.~Xu, and B.~Pfeiffer,  Chinese Physics C \textbf{36}, 1287 (2012).

\bibitem{cgp09} N.~Chamel, S.~Goriely, and J.~M.~Pearson,
Phys.\ Rev. C \textbf{80}, 065804 (2009).

\bibitem{gcp09}S.~Goriely, N.~Chamel, and J.~M.~Pearson,
Eur. Phys. Journ. A {\bf 42}, 547 (2009).

\bibitem{cha10} N.~Chamel, Phys.\ Rev. C \textbf{82}, 014313 (2010). 

\bibitem{pea20}J. M. Pearson, N. Chamel, and A. Y. Potekhin, 
Phys.\ Rev. C \textbf{101}, 015802 (2020).	

\bibitem{pea22}J. M. Pearson and N. Chamel,
Phys.\ Rev. C \textbf{105}, 015803 (2022).

\bibitem{shch23}N. N. Shchechilin, N. Chamel and J. M. Pearson,
Phys.\ Rev. C \textbf{108}, 025805 (2023).

\bibitem{rav} D.~G. Ravenhall, C.~J. Pethick, and J.~R. Wilson,
Phys.\ Rev.\ Lett. \textbf{50}, 2066 (1983).

\bibitem{hash}M.~Hashimoto, H.~Seki, and M.~Yamada,
Prog.\ Theor.\ Phys. \textbf{71}, 320 (1984).

\bibitem{pot98} A. Y. Potekhin and C. J. Pethick, 
Phys. Lett. \textbf{B427}, 7 (1998).

\bibitem{oy94}K.~Oyamatsu and M.~Yamada, Nucl.\ Phys. \textbf{A578}, 181
(1994).

\bibitem{ch07} N.~Chamel, S.~Naimi, E.~Khan, and J.~Margueron, 
Phys.\ Rev. C \textbf{75},  055806 (2007).

\bibitem{grill}F. Grill, J. Margueron, and N. Sandulescu,
Phys.\ Rev. C \textbf{84}, 065801  (2011).

\bibitem{baldo05} M. Baldo, U. Lombardo, E.E. Saperstein, S.V. Tolokonnikov, 
Nucl. Phys. {\bf A 750}, 409 (2005).

\bibitem{baldo07} M. Baldo, E.E. Saperstein, S.V. Tolokonnikov, 
Phys. Rev. C 76, 025803 (2007). 

\bibitem{baldo06} M. Baldo, E.E. Saperstein, S.V. Tolokonnikov, 
Nucl. Phys. {\bf A 775}, 235 (2006).

\bibitem{sp20}M. Shelley and A. Pastore, Universe \textbf{6}, 206 (2020).

\bibitem{sp21} M. Shelley and A. Pastore, Phys. Rev. C {\bf 103}, 035807 (2021). 

\bibitem{alc21} V. Allard and N. Chamel, Universe {\bf 7} (12), 470 (2021).  

\bibitem{cao06} L.G. Cao, U. Lombardo, and P. Schuck, Phys. Rev. C  \textbf{74}, 064301  (2006).

\bibitem{gcp16}S.~Goriely, N.~Chamel, and J.~M.~Pearson,
Phys.\ Rev. C \textbf{93}, 034337  (2016).

\bibitem{pcpg15} J.~M.~Pearson, N.~Chamel, A.~Pastore, and S.~Goriely,
Phys.\ Rev. C \textbf{91}, 018801 (2015).

\bibitem{ons08}M.~Onsi, A.~K.~Dutta, H.~Chatri, S.~Goriely, N.~Chamel, and
J.~M.~Pearson, Phys.\ Rev. C \textbf{77}, 065805 (2008).

\bibitem{ws33}E. Wigner and F. Seitz, 
Phys.\ Rev. \textbf{43}, 804  (1933). 

\bibitem{shch24}N. N. Shchechilin, N. Chamel, J. M. Pearson,
A. I. Chugunov, and A. Y. Potekhin, Phys. Rev. C {\bf 109}, 055802 (2024).

\bibitem{cgpo10} N. Chamel, S. Goriely, J. M. Pearson, and M. Onsi, 
Phys. Rev. C {\bf 81}, 045804 (2010). 

\bibitem{pizzo77}P. M. Pizzochero, L. Viverit, and R. A. Broglia,
Phys. Rev. Lett. {\bf 79}, 3347 (1997).

\bibitem{pcgd12} J.~M.~Pearson, N.~Chamel, S.~Goriely, and C. Ducoin,
Phys.\ Rev. C \textbf{85}, 065803  (2012).

\bibitem{cgp08} N. Chamel, S. Goriely, J.~M. Pearson, 
Nucl. Phys. {\bf A 812}, 72 (2008). 

\bibitem{gcp13b} S. Goriely, N. Chamel, J.~M.~Pearson, 
Phys. Rev. C {\bf 88}, 061302(R) (2013). 

\bibitem{carr2020a} T. Carreau, F. Gulminelli, N. Chamel, A. F. Fantina, J. M. Pearson, 
A\& A \textbf{635}, A84 (2020). 

\bibitem{carr2020b} T. Carreau, A. F. Fantina, F. Gulminelli, 
A\& A \textbf{640}, A77 (2020). 

\bibitem{pas17}  A. Pastore, M. Shelley, S. Baroni and C. A. Diget, 
J. Phys. G {\bf 44}, 094003 (2017). 

\bibitem{zho22} S. Zhou, E. G\^ugercino\u{g}lu, J. Yuan, M. Ge, C. Yu, Universe \textbf{8}, 641 (2022).

\bibitem{pear91} J.~M. Pearson, Y. Aboussir, A.~K. Dutta, R. C. Nayak, M. Farine,
and F. Tondeur, 
Nucl. Phys. {\bf A 528}, 1 (1991).

\bibitem{rs80}P. Ring, P. Schuck, {\it The Nuclear Many-Body Problem}, 
Springer, New York, 1980.

\bibitem{bbd}J. Bartel, M. Brack, and M. Durand, Nucl. Phys. {\bf A445},
263 (1985).

\bibitem{opp97}M. Onsi, H. Przysiezniak and J. M. Pearson, 
Phys. Rev. C {\bf 55}, 3139 (1997).

\bibitem{bgh85} M. Brack, C. Guet, and H.~B.
  H{\r{a}}kansson 
Phys. Rev. C \textbf{123},
  {275} ({1985}).

\end{thebibliography}
\end{document}